\documentclass[longauth]{aa}  
\usepackage{graphicx}
\usepackage{txfonts}
\defcitealias{assef22}{A22}
\defcitealias{assef25}{A25}
\defcitealias{mathis77}{MRN77}

\begin{document} 

   \title{Scattering from Dusty Outflows in Hot Dust-Obscured Galaxies with Blue Excess Emission}


   \author{
        R.~J. Assef\inst{\ref{inst1}}\and
        M. Stalevski\inst{\ref{inst2},\ref{inst3}} \and
        F.~E. Bauer\inst{\ref{inst4}} \and
        A. Blain\inst{\ref{inst5}} \and
        T. D\'iaz-Santos\inst{\ref{inst6},\ref{inst7}} \and
        P.~R.~M. Eisenhardt\inst{\ref{inst8}} \and
        R. Fern\'andez-Aranda\inst{\ref{inst9}} \and
        H.~D. Jun\inst{\ref{inst10},\ref{inst11}} \and
        M. Liao\inst{\ref{inst1}} \and
        E. Shablovinskaia\inst{\ref{inst12}} \and
        A. Sarath\inst{\ref{inst1}} \and
        M. Stepney\inst{\ref{inst13}} \and
        D. Stern\inst{\ref{inst8}} \and
        C.-W. Tsai\inst{\ref{inst14}, \ref{inst15}, \ref{inst16}} \and
        A. Vayner\inst{\ref{inst17}} \and
        D.~J. Walton\inst{\ref{inst18}} \and
        D. Wylezalek\inst{\ref{inst19}} \and
        D. Zewdie\inst{\ref{inst20}}
    }

   \institute{
        Instituto de Estudios Astrof\'isicos, Facultad de Ingenier\'ia y Ciencias, Universidad Diego Portales, Av. Ej\'ercito Libertador 441, Santiago, Chile.~\email{roberto.assef@mail.udp.cl}\label{inst1}
        \and
        Astronomical Observatory, Volgina 7, 11060 Belgrade, Serbia\label{inst2}
        \and
        Department of Physics and Astronomy, Universiteit Gent, Proeftuinstraat 86 N3, B-9000 Ghent, Belgium\label{inst3}
        \and
        Instituto de Alta Investigaci{\'{o}}n, Universidad de Tarapac{\'{a}}, Casilla 7D, Arica, Chile\label{inst4}
        \and
        Department of Physics and Astronomy, University of Leicester, University Road, Leicester LE1 7RH, UK\label{inst5}
        \and
        Institute of Astrophysics, Foundation for Research and Technology-Hellas (FORTH), Heraklion, 70013, Greece\label{inst6}
        \and
        School of Sciences, European University Cyprus, Diogenes street, Engomi, 1516 Nicosia, Cyprus\label{inst7}
        \and
        Jet Propulsion Laboratory, California Institute of Technology, Pasadena, CA 91109, USA\label{inst8}
        \and
        Centro de Astrobiolog\'ia (CAB), CSIC-INTA, Carretera de Ajalvir km 4, Torrej\'on de Ardoz, E-28850, Madrid, Spain\label{inst9}
        \and
        Department of Physics, Northwestern College, 101 7th St SW, Orange City, IA 51041, USA\label{inst10}
        \and
        School of Physics, Korea Institute for Advanced Study, 85 Hoegiro, Dongdaemun-gu, Seoul 02455, Republic of Korea\label{inst11}
        \and
        Max Planck Institut f\"ur Radio Astronomie, Auf dem H\"ugel 69, D-53121, Bonn, Germany\label{inst12}
        \and
        Centro de Astrof\'isica y Tecnologias Afines, Av. Vicu\~na Mackenna 4860, San Joaqu\'in, Santiago, Chile\label{inst13}
        \and
        Key Laboratory of Radio Astronomy and Technology, Chinese Academy of Sciences, Beijing 100101, China\label{inst14}
        \and
        School of Astronomy and Space Science, University of Chinese Academy of Sciences, Beijing 100049, China\label{inst15}
        \and
        Institute for Frontiers in Astronomy and Astrophysics, Beijing Normal University, Beijing 102206, China\label{inst16}
        \and
        Florida Gulf Coast University, 10501 FGCU Blvd. South, Fort Myers, 33965, FL, USA \label{inst17}
        \and
        Centre for Astrophysics Research, University of Hertfordshire, College Lane, Hatfield AL10 9AB, UK\label{inst18}
        \and
        Zentrum f\"ur Astronomie der Universit\"at Heidelberg, Astronomisches Rechen-Institut, M\"onchhofstr, 12-14 69120 Heidelberg, Germany\label{inst19}
        \and
        Centre for Space Research, North-West University, Potchefstroom 2520, South Africa\label{inst20}  
   }

   \date{}

  \abstract
   {Hot Dust-Obscured Galaxies (Hot DOGs) are a population of hyper-luminous quasars heavily obscured by dust and gas. While most have rest UV/optical SEDs dominated by their host galaxy emission, a small fraction shows blue emission in excess of what is expected for star-formation. Detailed multi-wavelength studies from X-rays to the far-IR of these Blue Excess Hot DOGs (BHDs) have concluded the most likely origin of this blue excess is scattered light from the central engine. This scenario was confirmed by previous studies for the BHDs W0116--0505 and W0204--0506 which found highly polarized UV emission, a key signature of scattered light.}
   {In this article we investigate this scattered light scenario in three other BHDs and further constrain the properties of the scattering material in W0116--0505.}
   {We analyze new VLT/FORS2 imaging polarimetry observations in the $R_{\rm Special}$-band of three BHDs and additional observations of W0116--0505 in the $v_{\rm High}$- and $I_{Bessel}$-bands.}
   {We find spatially integrated polarization fractions in all three BHDs observed in the $R_{\rm Special}$-band ranging from about 6\% to 15\%. This confirms that their blue excess emission is due to AGN scattered light. We also find that the polarization fraction has a significant dependence with rest-frame wavelength in W0116--0505 and that it cannot be explained by Thomson scattering or by an AGN shining onto a smooth ISM. We find that a dusty bi-conical outflow acting as the scatterer can successfully explain the observations as long as the dust is dominated by graphites. This may be due to the high temperatures of the outflowing gas and the higher sublimation temperature of graphites compared to silicates.}
   {}


   \maketitle

\section{Introduction}\label{sec:intro}

In massive galaxies, the feedback into the interstellar medium (ISM) from the energy released by gas accretion onto their central super-massive black holes (SMBHs) during Active Galactic Nucleus or AGN phases is thought to be a major component in the regulation of their stellar-mass growth \citep[e.g.,][]{dimatteo05}. This co-evolution between the SMBH and host galaxy is likely required to set empirical relations observed at low redshifts, such as those between the SMBH's mass, $M_{\rm BH}$, and the stellar mass, luminosity and velocity dispersion of the galaxy's spheroidal component \citep[e.g.][]{magorrian98, ferrarese00, marconi03}. Crucially, AGN feedback is also needed in cosmological simulations to avoid overpredicting the space density of high-stellar-mass galaxies in the local universe \citep[e.g., see][]{harrison17}. At least some, if not most, of this AGN feedback is thought to come from the interaction between the ISM and radiatively driven winds launched by the accretion disk. As such, the most striking effects of AGN feedback should be preferentially evident in the most luminous AGNs in the Universe, which we call quasars. 

Luminous quasar activity is thought to be related to merger activity \citep[e.g.,][]{treister12}, as the torque exerted on the ISM allows gas and dust to be efficiently funneled into the inner regions of the galaxy. The merger triggers dust-obscured star-formation throughout the galaxy, while the sudden gas and dust inflow into the inner regions eventually triggers an AGN that is also highly obscured by dust. The forming quasar should eventually become powerful enough to start clearing away some of this dust through winds and that, through the aforementioned AGN feedback, quench the star-formation of the host galaxy \citep[e.g.,][]{hopkins08, alexander12}. This picture implies that the most intense episodes of AGN feedback should occur in these highly obscured, high-accretion-rate stages of quasar activity. 

The most luminous, obscured quasars are, however, very rare, so identifying them was a difficult task until the advent of wide-field deep infrared surveys over the past two decades, such as the VISTA Hemisphere survey \citep[VHS;][]{vhsdr5} which mapped the southern sky in up to four NIR bands ($Y$, $J$, $H$ and $Ks$) and, most notably, the Wide-field Infrared Survey Explorer \citep[WISE;][]{wright10}, which observed the entire sky in four mid-IR bands dubbed W1 (3.4$\mu$m), W2 (4.6$\mu$m), W3 (12$\mu$m) and W4 (22$\mu$m). These surveys led to the discovery of complementary populations of hyper-luminous (i.e., $L_{\rm Bol}\gtrsim 10^{13}~L_{\odot}$) obscured quasars, such as heavily reddened type 1 quasars \citep[HRQs;][]{banerji15} selected through VHS observations, extremely red quasars \citep[ERQs;][]{ross15,hamann17} selected through a combination of Sloan Digital Sky Survey \citep[SDSS;][]{sdssdr17} and WISE observations, and Hot Dust-Obscured Galaxies \citep[Hot DOGs;][]{eisenhardt12,wu12}, selected solely using WISE observations. Due to their selection function requiring very red WISE colors \citep{eisenhardt12}, Hot DOGs represent the most obscured of these populations with $2\lesssim E(B-V)\lesssim 25$ \citep{assef15}, while HRQs, for example, have $0.5\lesssim E(B-V)\lesssim 2$ \citep{banerji15, stepney26a}. Hot DOGs are also among the most luminous of these populations \citep{stepney26b} with about 10\% of them exceeding $10^{14}~L_{\odot}$ \citep{tsai15}. As expected, Hot DOGs have been shown to have powerful ionized gas outflows through observations of their [O{\sc iii}]$\lambda 5007\rm \AA$ emission lines \citep{jun20,finnerty20,vayner25}. This is also true for ERQs \citep{zakamska16, perrotta19, vayner24} and HRQs \citep{temple19} with an interesting variety in their intensities \citep[see, e.g., the discussion by][hereafter \citetalias{assef22}]{assef22}.  

Owing to their large dust obscuration, the spectral energy distribution (SED) of Hot DOGs is typically dominated by the host galaxy emission at UV and optical wavelengths \citep{assef15}. However, a fraction of them \citep[which could be as high as 25\%, ][]{li24} show UV emission that is in excess of what would be expected for a star-forming galaxy. Furthermore, this excess emission is well modeled in their broad-band SEDs by adding an unobscured or mildly-obscured AGN component to the SED models that has $\sim1\%$ of the luminosity of the highly obscured, hyper-luminous quasar that powers the mid-IR \citep{assef16,assef20,li24}. While the additional emission in these blue excess Hot DOGs (or BHDs) could be due to a second AGN in the system or to an extreme, unobscured star-formation event, studies of their X-ray through IR SEDs and spectral properties concluded that the most likely scenario is that this excess emission is dominated by scattered light from the accretion disk \citep{assef16,assef20}. 

One of the main predictions of this scattered light scenario is that their UV continua should be highly polarized, and this has been proven to be the case for two BHDs: WISE J011601.41--050504.0 (W0116--0505, $z=3.173$) and WISE J020446.12--050640.8 (W0204--0506, $z=2.099$). For W0116--0505, \citetalias{assef22} obtained polarization imaging observations in the $R_{\rm Special}$ band using the FORS2 instrument at the VLT observatory and measured a polarization fraction of $p=10.8\pm 1.9\%$. They proposed that the scattering material could correspond to dust in a polar outflow, considering the powerful ionized-gas outflow detected in this object by \citet{finnerty20} through the kinematics of the [O{\sc iii}] emission line. However they found that the observations could be equally well fit if the AGN was shining into a smooth distribution of free electrons or dust in the ISM. For W0204--0506, \citet[][hereafter \citetalias{assef25}]{assef25} used the same observational setup as \citetalias{assef22} and found a spatially integrated polarization fraction of $p=24.7\pm 0.7\%$. Furthermore, they found the polarized emission to be spatially resolved and with a gradient from about 15\% to 35\% aligned with the extension of the UV emission observed in {\it HST}/WFC3 F555W. They concluded that the extension corresponds to a massive outflow highly inclined from the line of sight, and that dust in this outflow acts as the scattering material, consistent with the scenario proposed by \citetalias{assef22} for W0116--0505. Furthermore, using radiative transfer simulations and assuming a conic polar outflow geometry, they found that the dust responsible for the scattering must be graphite-rich, as dust mixtures with significant amounts of silicates cannot reproduce the observed high polarization fraction. 

Scattered light from the central engine is a common occurrence in hyper-luminous, highly obscured AGN, having found to be nearly ubiquitous in ERQs \citep{alexandroff18,zakamska23} and HRQs \citep{stepney24,stepney26b}. The much lower fraction of Hot DOGs that show scattered light is consistent with an evolutionary picture where Hot DOGs probe the initial, most dust-obscured and intense stages of SMBH accretion, while ERQs and HRQs probe stages where the dust blow-out is already starting to clear significant lines of sight to the quasar. This is also suggested by the much lower central engine dust obscurations of HRQs as compared to Hot DOGs. Furthermore, \citet{stepney26b} finds that HRQs have an order of magnitude lower scattered-light fractions than BHDs, as HRQs scatter $\sim0.1\%$ of the incident light from the accretion disk while BHDs scatter closer to $\sim1\%$ \citep{assef16,li24}. This could indicate a lower column density of scattering material in HRQs as compared to BHDs, consistent with this evolutionary scenario, although it could also be related to different inclinations with respect to the line of sight \citep[see discussion in][]{stepney26b}.

In this article we present new imaging polarimetric observations of W0116--0505 in the $I_{\rm Bessel}$ and $v_{\rm High}$ broad-bands to complement the previous observations by \citetalias{assef22} in the $R_{\rm Special}$ broad-band. Additionally, we present new imaging polarimetric observations in the $R_{\rm Special}$ broad-band of three additional BHDs identified by \citet{assef16}, and reanalyze those presented by \citetalias{assef25} for W0204--0506. In Section \ref{sec:obs} we discuss the imaging polarimetric observations as well as the supporting broad-band photometry and spectroscopy used in this study. In Section \ref{sec:pol_measurements} we describe the methodology used to measure the polarization fraction and angle, and present spatially integrated and resolved measurements. In Section \ref{sec:discussion} we discuss the constraints on the properties of the scattering material using different models, and in Section \ref{sec:conclusions} we summarize our conclusions. We assume a vanilla $\Lambda$CDM cosmology with $H_0=70~\rm km~\rm s^{-1}~\rm Mpc^{-1}$ and $\Omega_{\rm M} = 0.3$ whenever needed. 

\section{Observations}\label{sec:obs}

\subsection{Imaging Polarimetry}\label{ssec:impol_data}

We have obtained new imaging polarimetric observations in the $R_{\rm Special}$ broad-band using the FORS2 instrument at the VLT observatory (Program ID 111.24UL) for a sample of three BHDs identified by \citet{assef16}: WISE J001926.87--104633.2 (W0019--1046, $z=1.641$), WISE J022052.12+013711.5 (W0220+0137, $z=3.122$) and WISE J083153.24+014010.7 (W0831+0140, $z=3.888$). The same observing program also obtained the $R_{\rm Special}$ imaging polarization observations of W0204--0506 presented by \citetalias{assef25}, which we incorporate into our present analysis for completeness. We also obtained observations of W0116--0505 in the $v_{\rm High}$ and $I_{\rm Bessel}$ bands, which had previously been observed in the $R_{\rm Special}$ band by \citetalias{assef22}. We make extensive use of the observations by \citetalias{assef22} throughout this article and re-analyze them here for consistency with the rest of the targets. Our targets are summarized in Table \ref{tab:targs}. 

\begin{table*}
    \renewcommand{\arraystretch}{1.5}
    \caption{\label{tab:targs} BHD Targets}
    \scriptsize
    \begin{tabular}{lcccccccccccc}
        \hline \hline
        &  &  &  & &  \multicolumn{2}{c}{Obscured AGN} & \multicolumn{2}{c}{Scattered AGN} & & & & \\
        WISE ID & R.A. & Dec. & $r$ & $z$ & $\log{L_{6\mu\rm m}}$ & $E(B-V)$ & $\log{L_{6\mu\rm m}}$ & $E(B-V)$ & $P_{\rm ran}$ & $\log{M_{\rm BH}}$ & $\log{L_{\rm Bol}}$ & $\lambda_{\rm Edd}$\\
         & (J2000) & (J200) & (mag) & & ($\rm erg~\rm s^{-1}$) & (mag) & ($\rm erg~\rm s^{-1}$) & (mag) & ($10^{-2}$) & ($M_{\odot}$) & ($L_{\odot}$) & \\
        \hline
        W0019$-$1046 & 00:19:26.88 & $-$10:46:33.3 & 22.0 & 1.641 & $46.53^{+0.04}_{-0.03}$ & $ \phantom{0}5.01^{+0.42}_{-0.14}$ & $44.58^{+0.23}_{-0.14}$ & $ 0.03^{+0.05}_{-0.03}$ & 5.071 & 9.8 & 13.3 & 0.10$^{+0.17}_{-0.06}$ \\
        W0116$-$0505 & 01:16:01.41 & $-$05:05:04.1 & 21.4 & 3.173 & $47.28^{+0.11}_{-0.08}$ & $\phantom{0}4.32^{+1.47}_{-0.86}$ & $45.22^{+0.02}_{-0.05}$ & $ 0.00^{+0.01}_{-0.00}$ & 0.021 & 9.4 & 14.1 & 1.53$^{+2.75}_{-0.98}$\\
        W0204$-$0506 & 02:04:46.13 & $-$05:06:40.8 & 22.5 & 2.099 & $46.89^{+0.04}_{-0.11}$ & $10.00^{+0.21}_{-2.06}$ & $45.00^{+0.05}_{-0.49}$ & $ 0.10^{+0.01}_{-0.08}$ & 9.743 & 8.8 & 13.7 & 2.42$^{+4.36}_{-1.55}$\\
        W0220+0137 & 02:20:52.12 & +01:37:11.6 & 21.8 & 3.122 & $47.37^{+0.05}_{-0.06}$ & $ \phantom{0}7.57^{+0.84}_{-1.02}$ & $45.07^{+0.01}_{-0.02}$ & $ 0.00^{+0.01}_{-0.00}$ & 0.001 & 9.3 & 13.9 & 1.21$^{+2.18}_{-0.78}$\\
        W0831+0140 & 08:31:53.25 & +01:40:10.8 & 21.8 & 3.888 & $47.54^{+0.04}_{-0.02}$ & $ \phantom{0}3.16^{+0.48}_{-0.12}$ & $45.26^{+0.19}_{-0.14}$ & $ 0.02^{+0.02}_{-0.02}$ & 0.387 & 9.4 & 14.4 & 3.04$^{+5.48}_{-1.96}$\\
        \hline
    \end{tabular}
    \tablefoot{The uncertainties for $M_{\rm BH}$ and $L_{\rm Bol}$ are 0.4~dex and 0.2~dex respectively, dominated by the systematic uncertainties of the methods used. See \citet{li24} for details.}
\end{table*}

Observations were obtained using retarder plate angles of 0, 22.5, 45, and 67.5 degrees, suggested in the FORS2  Manual\footnote{\url{https://www.eso.org/sci/facilities/paranal/instruments/fors/doc.html}} as the minimum advisable that must be used to suppress the effects of improper flat-fielding of the data. Observations were divided into two observing blocks (OBs) in the $R_{\rm Special}$ and $I_{\rm Bessel}$ bands, while a single OB was used in the $v_{\rm High}$ band. We obtained 2$\times$353s exposures at each retarder plate angle in each of the $R_{\rm Special}$ and $v_{\rm High}$ band OBs, and 3$\times$228s exposures in each $I_{\rm Bessel}$ band OB. Observations using the same set of retarder plate angles were obtained as part of the nighttime calibrations program (PID 60.A–9203(E)) for a number of polarization and zero-polarization standard stars, and we use those closest in time to our observations that were not saturated nor out of focus to avoid systematic uncertainties in the measurements. A summary of the observations is presented in Table \ref{tab:data}.

We applied bias subtraction to all images using the {\tt{EsoRex}}\footnote{\url{https://www.eso.org/sci/software/cpl/esorex.html}} pipeline. Following \citetalias{assef22}, we do not apply flat-field or polarization flat-field corrections to our observations, and remove cosmic rays using the {\tt{Python}} package {\tt{Astro-SCRAPPY}} \citep{astroscrappy}, which is based on the algorithm of \citet{vandokkum01}.

\subsection{Optical Spectroscopy}

\begin{figure}
    \centering
    \includegraphics[width=0.49\textwidth]{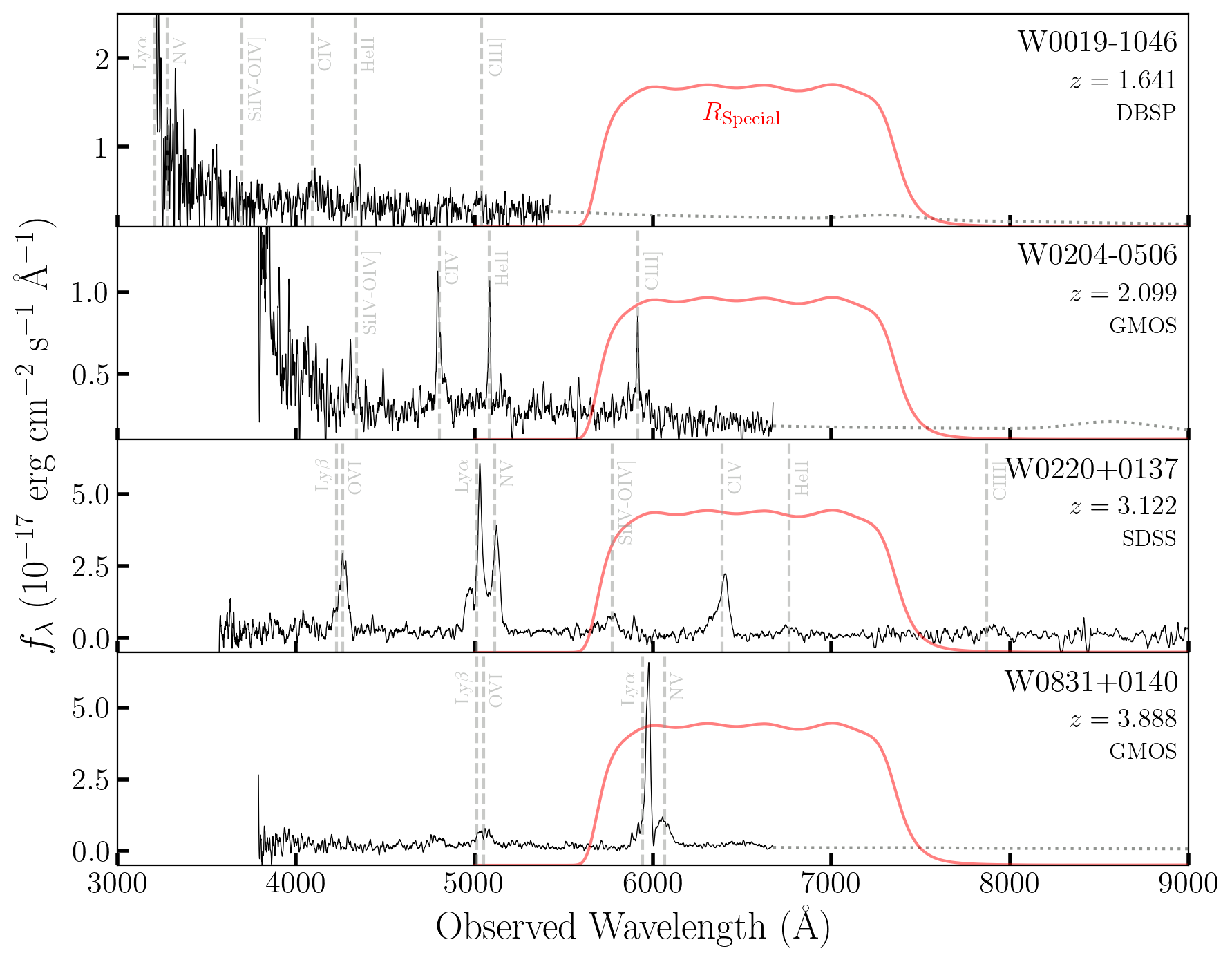}
    \caption{Spectra of all four BHDs studied in this article for which only $R_{\rm Special}$ observations were obtained. Each panel also shows the broad-bands for which we have imaging polarimetry observations (see \S\ref{ssec:impol_data}). The spectra have been extended using the best-fit SED model (dotted gray line) when needed to overlap with the polarimetry broad-bands.}
    \label{fg:specs}
\end{figure}

\begin{figure}
    \centering
    \includegraphics[width=0.49\textwidth]{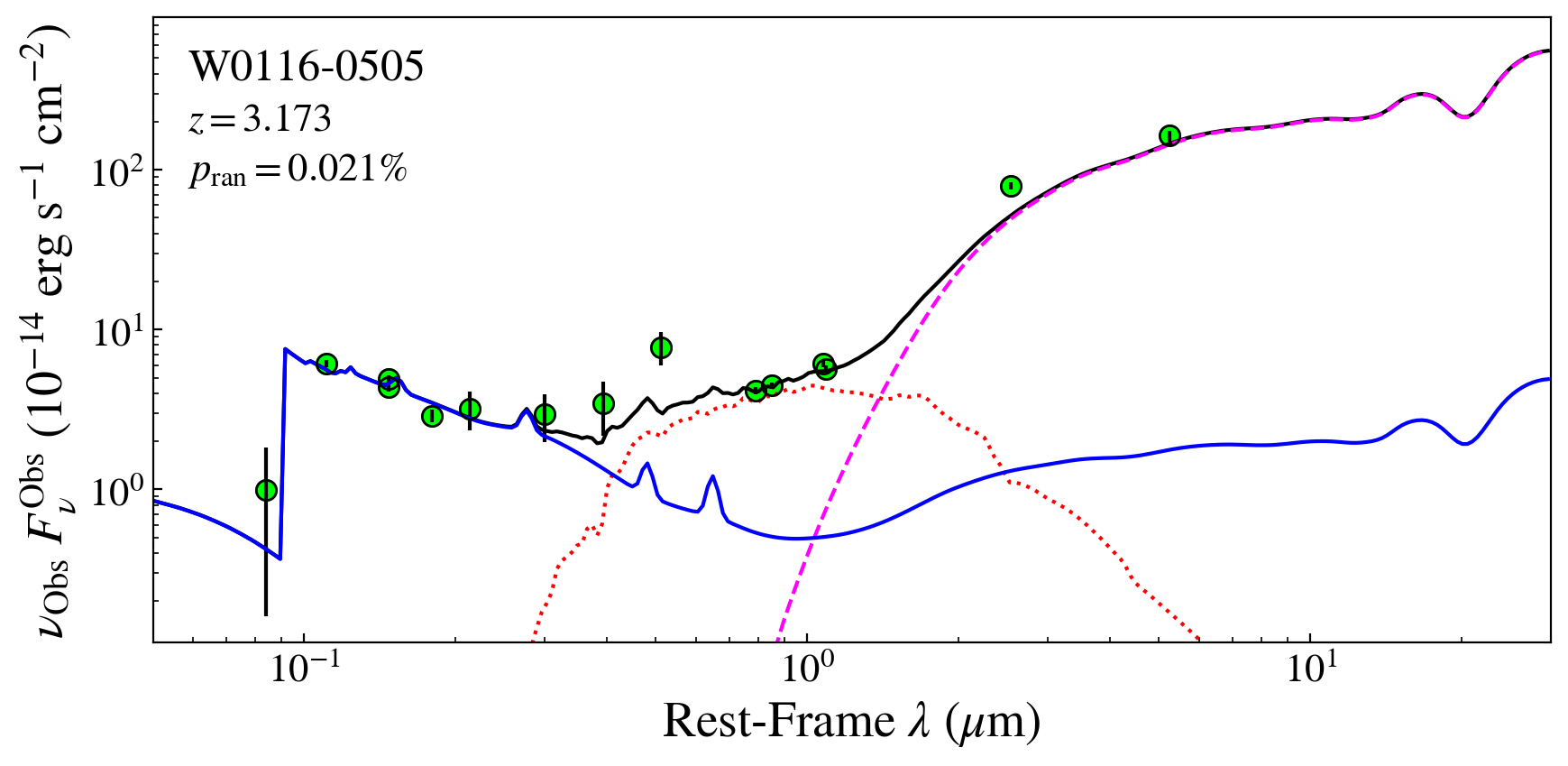}
    \includegraphics[width=0.49\textwidth]{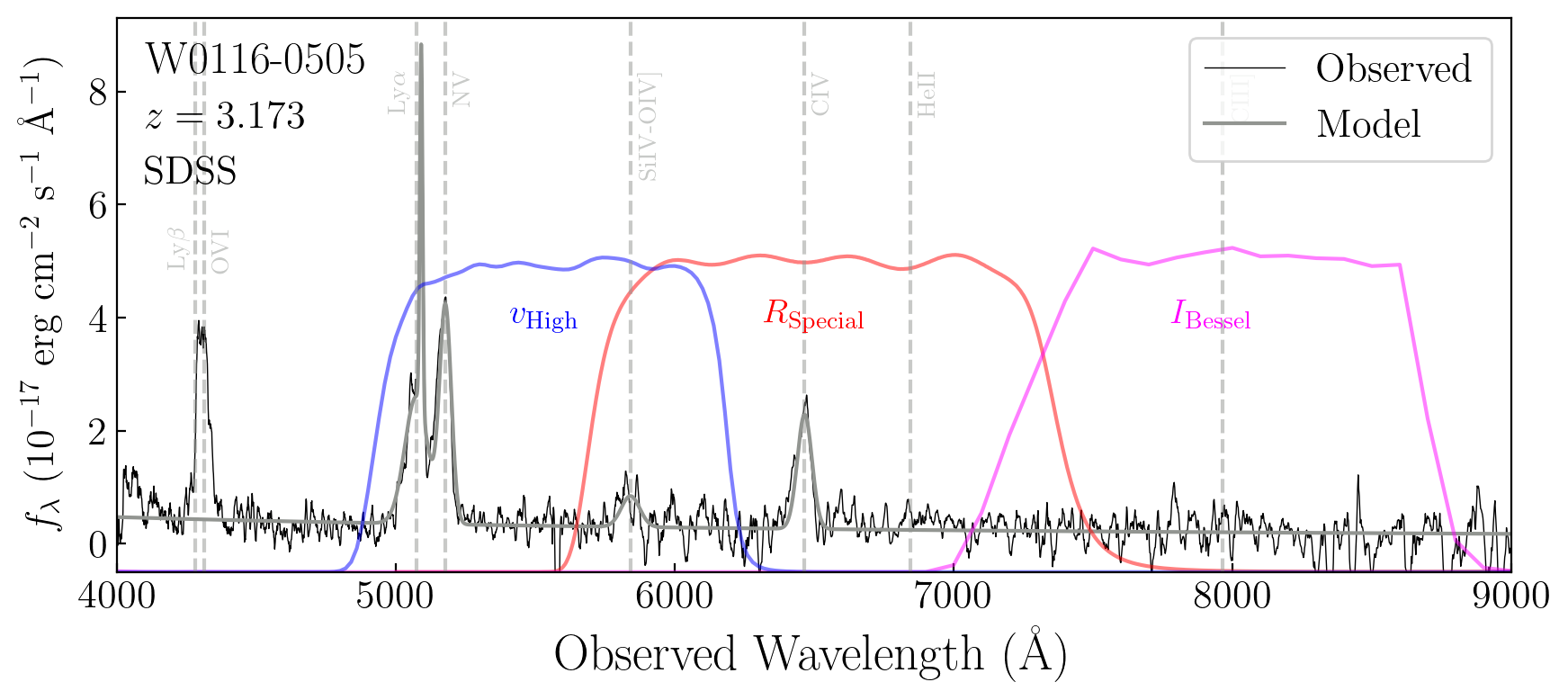}
    \caption{{\it{(Top panel)}} Broad-band SED of the BHD W0116--0505, for which we present new polarimetry observations in the $v_{\rm High}$ and $I_{\rm Bessel}$ bands. The solid black line shows the best-fit model, consisting of two AGN (obscured primary in dashed magenta and unobscured/lightly obscured secondary in solid blue) and the best-fit host galaxy component (E in dotted red) from the templates of \citet{assef10}. {\it{(Bottom panel)}} Spectrum of W0116--0505 obtained by SDSS (solid black line). The best-fit spectral model is shown by the solid gray line (see Appendix \ref{app:thomson_thin_scatt} for details).}
    \label{fg:SED_w0116}
\end{figure}

Spectroscopic observations of the five objects studied here are shown in Figures \ref{fg:specs} and \ref{fg:SED_w0116}. The spectra of W0116--0505 and W0220+0137 were obtained by SDSS and retrieved from its archive. Spectra for W0204--0506 and W0831+0140 were obtained using the GMOS-S spectrograph at the Gemini Observatory and were presented by \citet{assef20} and \citet{diaz-santos21}, respectively. We refer the reader to those references for details on these observations. The spectrum of W0019--1046 was obtained using the Double Spectrograph (DBSP) at the Palomar 200-inch telescope on UT 2012-07-14, using the D55 dichroic with the 300/3900 grating in the blue arm and the 316/7500 in the red arm, although the object was only detected with sufficient SNR to extract the 1D spectrum in the blue arm. Observations were obtained using a 1.5\arcsec\ slit with an exposure time of $6\times 600\rm s$, and were reduced using standard tasks from {\tt{IRAF}}\footnote{\url{https://iraf-community.github.io/}} following the procedure described in \citet{assef20}. Redshifts were measured from the centroid of the emission lines. A detailed characterization of the spectroscopic observations of Hot DOGs in general will be presented by \citet{eisenhardt24}.

Quasar broad emission lines are well detected in all the spectra except for that of W0019--1046, where, other than Ly$\alpha$, there is only a 4$\sigma$ detection of a broad C{\sc iv}$\lambda$1549\AA\ emission line that was judged to be credible based on the 2D spectrum. Recently, \citet{li24} led a study on BHDs and estimated their SMBH masses. Following the same procedure detailed there, we obtain SMBH masses in the range of $\log{M_{\rm BH}/M_{\odot}} = 8.8-9.8$ with uncertainties of 0.4~dex dominated by the systematic uncertainties of the method (see Table \ref{tab:targs}). Their bolometric luminosities (see \citealt{tsai15} and \citealt{li24} for details) are in the range of $\log{L_{\rm Bol}/L_{\odot}} = 13.3-14.4$ with 0.2~dex uncertainty (see Table \ref{tab:targs}). These values imply super-Eddington accretion ratios, $L_{\rm Bol}/L_{\rm Edd} = \lambda_{\rm Edd}\ge 1$ where $L_{\rm Edd}$ is the Eddington luminosity, for all targets except for W0019--1046, which instead has a rather low accretion rate at $\lambda_{\rm Edd} = 0.10^{+0.17}_{-0.06}$ (see Table \ref{tab:targs}). This is likely due to the low SNR of the detection of the C{\sc iv} emission line used for the SMBH mass estimate in \citet{li24} which leads to a very broad FWHM estimate of over $6000~\rm km~\rm s^{-1}$, about 2.5--4 times larger than that of the other BHDs studied. As the SMBH mass estimates depend on the square of this quantity, such an overestimate of the line-width would result in a 6--16 times underestimated $\lambda_{\rm Edd}$.

\subsection{Photometry}\label{ssec:phot}

We construct the SEDs of the five imaging polarimetry targets using photometry from public surveys. Specifically, we use photometry from the Sloan Digital Sky Survey (SDSS) DR17 \citep{sdssdr17} in the $u$, $g$, $r$, $i$ and $z$ bands, from CatWISE2020 \citep{eisenhardt20} in the WISE W1 and W2 bands and from the WISE AllSky Data Release \citep{cutri12} in W3 and W4. Whenever available, we also add near-IR photometry in the $J$, $H$ and $Ks$ bands from the VHS DR5 \citep{vhsdr5} and mid-IR photometry from the {\it{Spitzer}}/IRAC [3.6] and [4.5] observations presented by \citet{griffith12}, as well as targeted $r$-band photometry obtained with SOAR/SOI \citep[see][for details]{assef20, eisenhardt24} and the $J$, $H$ and $Ks$ photometry presented by \citet{assef15}. We prefer using SDSS rather than Legacy Survey \citep{dey19} photometry despite the observations of the latter being generally deeper. This is because all five sources investigated here are classified as point sources in the Legacy Survey DR10 and only PSF-fitting photometry is provided despite them being marginally resolved. We note, however, that if we use the Legacy Survey photometry instead of SDSS, our results do not qualitatively change. 

\begin{figure}
    \centering
    \includegraphics[width=0.49\textwidth]{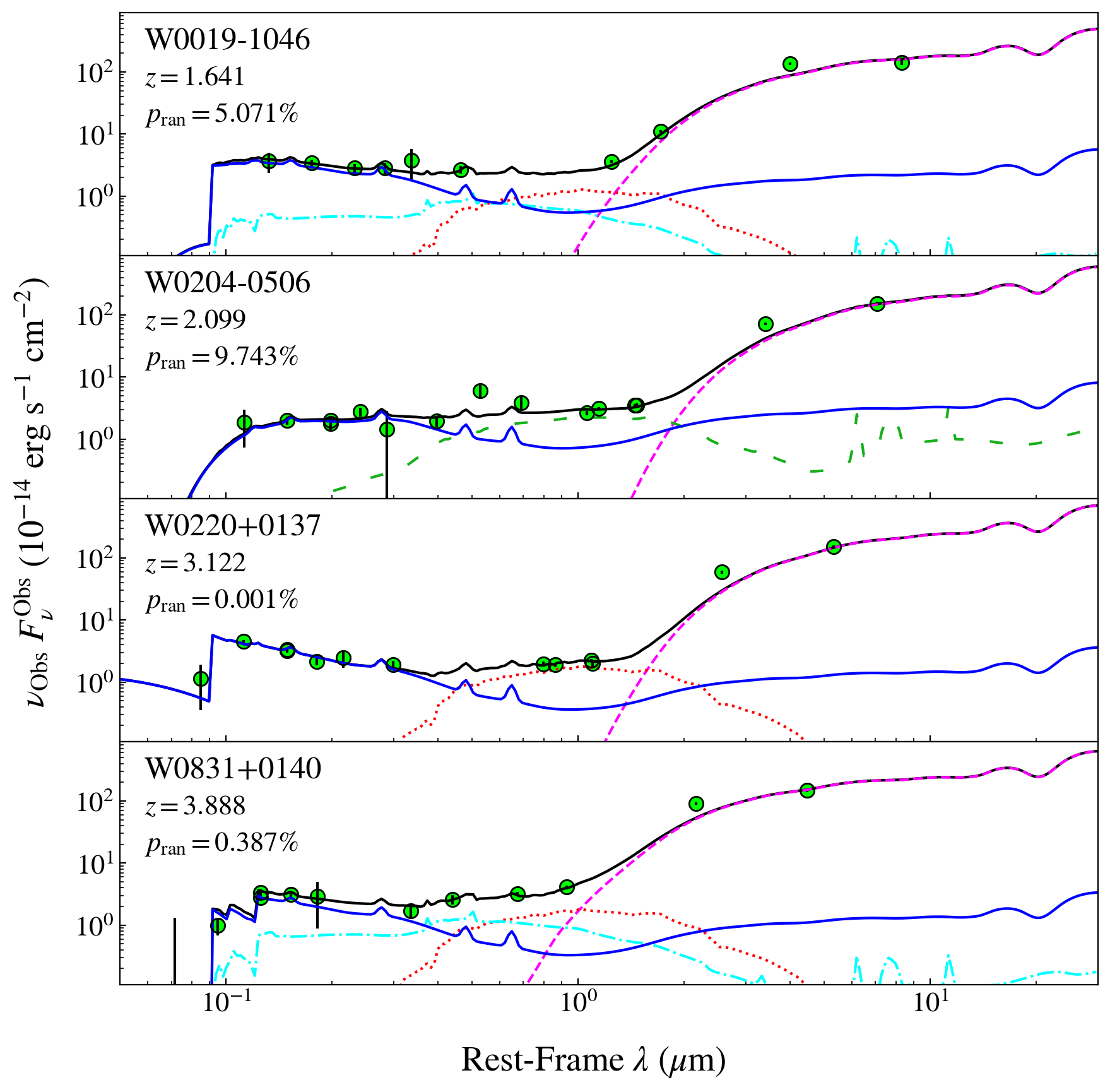}
    \caption{SEDs of the four BHDs with $R_{\rm Special}$ polarimetry observations discussed in this article. The green points show the photometry described in \S\ref{ssec:phot}. The solid black line shows the best-fit model, consisting of two AGN (obscured primary in dashed magenta and unobscured/lightly obscured secondary in solid blue) and three host galaxy components (E in dotted red, Sbc in dashed green and Im in dot-dashed cyan) from the templates of \citet{assef10}.}
    \label{fg:SEDs}
\end{figure}

For consistency, we model the SEDs following the same approach of \citet{assef16}. Specifically, we use the algorithm of \citet{assef10} that models the SED of an object as a non-negative combination of four empirically derived SED templates consisting of one AGN and three galaxy templates, where one resembles an elliptical (E), one resembles an intermediate spiral (Sbc) and the other resembles a local starburst (Im). Reddening is also fit for the AGN component and it is parameterized by $E(B-V)$, while the IGM absorption strength is fit for the combination of all templates. We then modify this approach, as done by \citet{assef16}, to add a second AGN component with an independent amount of dust reddening. \citet{assef16} defined BHDs as objects for which the second AGN component to model their SEDs is justified based on an F-test \citep[see also][]{li24}. Figures \ref{fg:SED_w0116} and \ref{fg:SEDs} show the SEDs of the five BHDs studied in this article along with the best-fit model using two AGN templates. Each panel also indicates the value of $p_{\rm ran}$, the probability that the improvement in $\chi^2$ from the addition of the second AGN component is spurious. The luminosity and reddening of both AGN components as well as the $p_{\rm ran}$ values are listed in Table \ref{tab:targs}. The best-fit SED model for W0204--0506 is also presented by \citetalias{assef25}, where they point out that the excess flux observed in the $H$-band may be consistent with a large outflow of ionized gas detected in [O{\sc iii}], a common feature among Hot DOGs \citep[see, e.g.,][]{jun20, finnerty20, vayner25}. \citetalias{assef25} also point out that removing this photometric band does not significantly alter the best-fit SED, although we note that the probability $p_{\rm ran}$ does decrease significantly, from 9.7\% to 3.3\%, strengthening its BHD classification. A similar offset is observed in the $Ks$ band of W0116--0505 (Fig. \ref{fg:SED_w0116}), which, as mentioned earlier, is known to have a large, fast ionized gas outflow detected in its [O{\sc iii}] emission line \citep{finnerty20}.

\section{Polarization Measurements}\label{sec:pol_measurements}

We measure the linear polarization fractions and angles following the procedure described by \citetalias{assef22}, which in turn is based on the approaches detailed by \citet{gg20} and in the FORS2 User Manual. Let $F$ be the normalized flux difference between the source fluxes in the ordinary ($f^o$) and extraordinary ($f^e$) beams, namely

\begin{equation}\label{eq:F_theta}
F(\theta_i) = \frac{f^o(\theta_i)-f^e(\theta_i)}{f^o(\theta_i)+f^e(\theta_i)} ,
\end{equation}

\noindent where $\theta_i$ is the angle of the retarder plate. The Stokes $Q$ and $U$ parameters are then estimated as

\begin{eqnarray}
Q &=& \sum_{i=1}^N \left(\frac{2}{N}\right) F(\theta_i) \cos{4\theta_i}\\
U &=& \sum_{i=1}^N \left(\frac{2}{N}\right) F(\theta_i) \sin{4\theta_i} ,
\end{eqnarray}

\noindent where $N=4$ is the number of retarder plate angles observed (see \S\ref{ssec:impol_data} for details). Note that in this definition, $Q$ and $U$ are already normalized by the Stokes parameter $I$. The linear polarization fraction is then given by 

\begin{equation}\label{eq:P}
P = \sqrt{Q^2+U^2} ,
\end{equation}

\noindent and the polarization angle by 

\begin{equation}\label{eq:chi}
\chi = \frac{1}{2} \arctan{\left(\frac{U}{Q}\right)} .
\end{equation}

We first study the spatially integrated measurements of the linear polarization fraction and the polarization angle for each source. Afterwards we attempt to look at the spatial distribution of these parameters in the sources that are spatially resolved in our observations. The spatially integrated measurements are the most robust and provide a more direct comparison to other sources in the literature, while the spatially resolved profiles can provide further insights into the geometry and nature of the scattering materials.    

\subsection{Spatially Integrated Measurements}\label{ssec:1d_pol}

Following \citetalias{assef22}, we use the {\tt{photutils}}\footnote{Version 1.11.0.} {\tt{Python}} package to measure the ordinary and extraordinary beam fluxes of each source through a circular aperture placed at the centroid of the emission in each beam of each frame. We use a 2\arcsec\ diameter aperture for the BHDs and either 2\arcsec\ or 3\arcsec\ diameter apertures for the standard stars depending on the quality of those observations. Before measuring the fluxes, the frames have been background subtracted and PSF-matched to the image with the worst seeing conditions. For the science target images we use the {\tt{ImageMatch}}\footnote{https://github.com/obscode/imagematch} software from Carnegie Observatories. For the standard stars, however, the images are significantly shallower which makes it difficult to properly match the PSFs, so we instead convolve to a common PSF assuming a 2D Gaussian shape using the {\tt{make\_2dgaussian\_kernel}} function of {\tt{photutils}} and the {\tt{convolve}} routine from {\tt{astropy}}. We note, however, that if we skip the PSF-matching steps and subtract the background locally from the source using annuli, we find consistent results. We then add the ordinary and extraordinary beam fluxes for all the frames obtained with the same retarder plate angle $\theta$, calculate the value of $F(\theta)$ as defined in equation (\ref{eq:F_theta}), and calculate the linear polarization fraction and the polarization angle using equations (\ref{eq:P}) and (\ref{eq:chi}). Uncertainties are measured using the Monte Carlo approach described in \citetalias{assef22}. A small error was discovered in the implementation of the Monte Carlo approach that led \citetalias{assef22} to overestimate the uncertainties, which we correct here. 

Table \ref{tab:pols_short} shows the linear polarization fractions and angles measured for each of our targets. In Table \ref{tab:pols} we show the measurements separated by OB as well as all the measurements for the standard stars. For the zero-polarization standards we see values that are generally consistent with no polarization, although in some cases a linear polarization is formally measured with an almost 2$\sigma$ significance. We do not identify any concerns for significant systematic issues through visual inspection of the images, with the possible exception of the $R_{\rm Special}$ observations of WD 2359--434 obtained on MJD 60293, which show high background levels. These marginal detections suggest that either our uncertainties are somewhat underestimated or that there may be a small level of systematic uncertainties, due to, for example, changes in the atmosphere between rotations, that we have not taken properly into account. We note, however, that these residual linear polarization fraction values are $\lesssim 0.3\%$, significantly smaller than any of the polarization fractions found in the targets of this study. 

\begin{table}
    \caption{\label{tab:pols_short} Polarization Measurements}
    \begin{tabular}{lccc}
        \hline \hline\\
        Target & Band &  P   & $\chi$ \\
               &      & (\%) &  (deg) \\
        \hline\\
        W0019--1046  & $R_{\rm Special}$ & \phantom{0}6.35$\pm$0.79 &           164.5$\pm$3.5 \smallskip\\
        W0116--0505  & $I_{\rm Bessel}$  &           14.36$\pm$0.43 & \phantom{0}72.7$\pm$0.9  \smallskip\\
                     & $R_{\rm Special}$ &           11.11$\pm$0.22 & \phantom{0}73.9$\pm$0.6  \smallskip\\
                     & $v_{\rm High}$    & \phantom{0}9.62$\pm$0.38 & \phantom{0}73.4$\pm$1.1  \smallskip\\
        W0204--0506  & $R_{\rm Special}$ &           24.72$\pm$0.66 & \phantom{0}12.7$\pm$0.8 \smallskip\\
        W0220+0137   & $R_{\rm Special}$ &           13.66$\pm$0.39 &           150.4$\pm$0.8 \smallskip\\
        W0831+0140   & $R_{\rm Special}$ & \phantom{0}7.45$\pm$0.40 &           131.5$\pm$1.6 \\
        \hline
    \end{tabular}
\end{table}        

We also measured the polarization fraction for three polarization standards found in the archive that were observed with the same filters within a span of a few weeks of our observations. While more were observed during the relevant time span, the majority were affected by systematic issues like saturation or improper focus. For BD-12 5133, our measurement of 4.13$\pm$0.04\% (4.17$\pm$0.06\%) for the linear polarization fraction and 146.5$\pm$0.2 deg (146.6$\pm$0.4 deg) for the polarization angle in the $R_{\rm Special}$ ($v_{\rm High}$) band are close to the values of 4.02$\pm$0.02\% (4.37$\pm$0.04\%) and 146.97$\pm$0.13 deg (146.84$\pm$0.25 deg) reported by \citet{fossati07} in the $R$ ($V$) band. For Vela-1, our measurements of 7.79$\pm$0.06\% (7.13$\pm$0.03\%) and 171.8$\pm$0.2 deg (171.6$\pm$0.1 deg) are close to the values of 7.89$\pm$0.04 (7.17$\pm$0.04\%) and 172.1$\pm$0.2 deg (172.2$\pm$0.2 deg) reported in the FORS homepage\footnote{\url{https://www.eso.org/sci/facilities/paranal/instruments/fors/inst/pola.html}} for the $R$ ($I$) band. The agreement for these polarization standards combined with the negligible linear polarization fraction found for the zero polarization standards earlier suggest that our measurements do not suffer from significant biases. 

Table \ref{tab:pols_short} shows that all BHDs have highly polarized rest UV emission. In the $R_{\rm Special}$ band, the five BHDs show linear polarization fractions ranging from $\sim$6\% in the case of W0019--1046 to $\sim$25\% in the case of W0204--0506, and are of a similar order of magnitude to the linear polarization fraction of about 11\% found for W0116--0505 by \citetalias{assef22} in the same broad band. These high polarization fractions confirm the conclusions of \citet{assef16} and \citet{assef20} that the excess blue emission in BHDs arises from scattered light of the central, highly obscured quasar.

In the $I_{\rm Bessel}$ observations of W0116--0505 we find a linear polarization fraction of 14.36$\pm$0.43\% which is significantly larger than the 11.11$\pm$0.22\% found in the $R_{\rm Special}$ band, which in turn is significantly larger than the 9.62$\pm$0.38\% found in the $v_{\rm High}$. Despite this change in polarization fraction with wavelength, the polarization angle $\chi$ is essentially the same for all 3 filters. We discuss this wavelength dependence in Section \ref{sec:discussion}. 

\subsection{Spatial Distribution of the Polarized Emission}\label{ssec:2d_pol}

\begin{figure*}
    \centering
    \includegraphics[width=\textwidth]{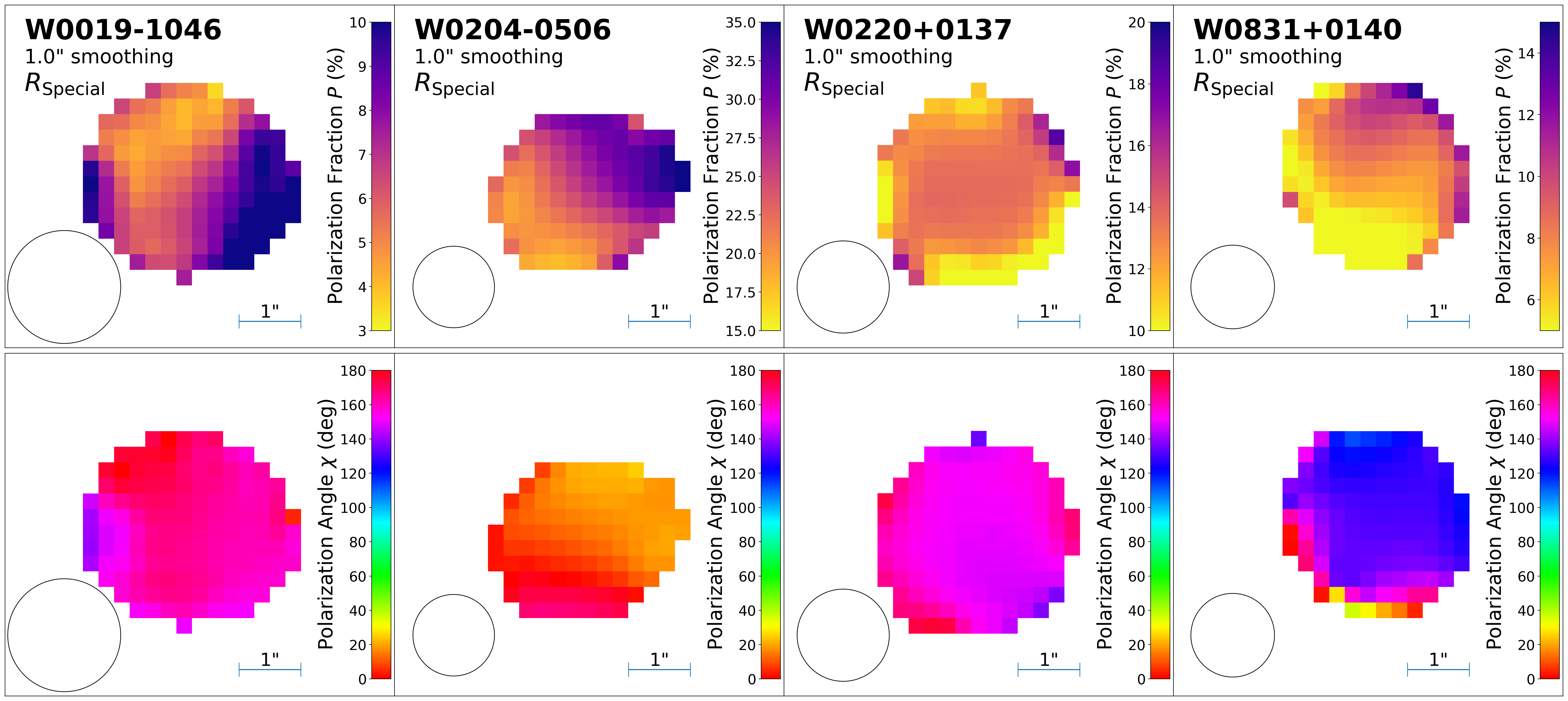}
    \caption{Spatial distribution of the polarization for the four BHDs for which we present observations in the $R_{\rm Special}$ band. Values are shown for every pixel for which flux in the stacked image of all of the o- and e-beam images is detected above 5$\sigma$. All o- and e-beam images where PSF-matched before combining (see text for details). An additional smoothing with a 1.0\arcsec\ kernel has been applied to all o- and e-beam images before combining. The circle on the bottom-left corner of each panel shows the effective size of the beam after smoothing. The top panels show the polarization fraction while the bottom panels show the polarization angle.}
    \label{fg:new_HDs_2Dpol}
\end{figure*}

Considering the spatially extended polarization found by \citetalias{assef25} for W0204--0506, we study the spatial properties of the polarization signal for the full sample of our BHDs. For any given target, differences in the PSF as well as small positional offsets are observed between images, even between those obtained within the same OB, so we start from the PSF-matched images discussed in the previous section. We note that {\tt{ImageMatch}} also aligns the images, so we can then simply compute the polarization fraction and angle for each pixel independently. Before doing the computation, however, we further apply smoothing using a Gaussian function with a 1.0\arcsec\ PSF to mitigate the noise. Figure \ref{fg:new_HDs_2Dpol} shows the spatial distribution of the polarization for the four BHDs for which we have observations only in the $R_{\rm Special}$ band, while Figure \ref{fg:W0116_2Dpol} shows it for W0116--0505 in the three bands we have observed. In Appendix \ref{app:2d_pol} we show the 2D polarization maps without applying the smoothing as well as dividing them by the specific OBs. They confirm that the patterns observed are not driven by a single set of observations, as large scale patterns in polarization maps can occur due to imperfect image registration, as well as from residual PSF or geometric distortions between the different retarder plate angles.

\begin{figure}
    \centering
    \includegraphics[width=0.5\textwidth]{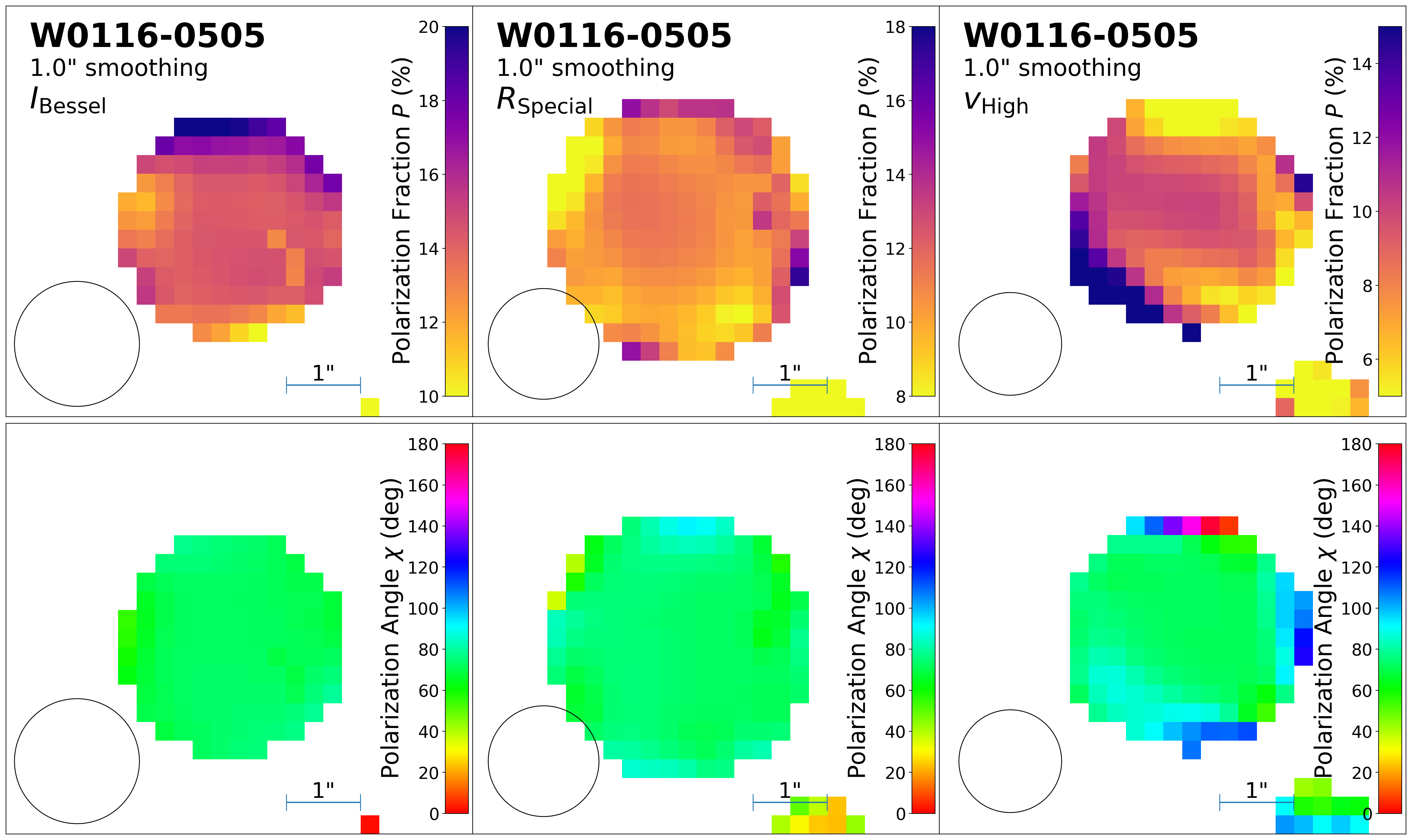}
    \caption{Same as Figure \ref{fg:new_HDs_2Dpol} but for W0116--0505 in the $I_{\rm Bessel}$, $R_{\rm Special}$ and $v_{\rm High}$ filters.}
    \label{fg:W0116_2Dpol}
\end{figure}

As reported by \citetalias{assef25}, W0204--0506 shows a significant gradient in the polarization fraction $P$ from about 15\% in the south east to about 35\% in the north west. The gradient is also observed in the polarization angle $\chi$, which shifts from about 0 to 20 degrees in the same direction. The implications of these gradients, particularly in comparison with the structures observed in {\it{HST}} imaging of the source have been previously discussed by \citetalias{assef25}. A similar pattern is also visible in W0019--1046, with a polarization fraction gradient from $\sim$4\% in the northeast to $\sim$10\% in the southwest, accompanied by a polarization angle change from about 180~deg to 150~deg in the same direction. Unfortunately no higher spatial resolution imaging for this source is currently available, but it is noteworthy that, along with W0204--0506, they are the two lowest redshift sources in our sample, for which spatial patterns should be easiest to detect due to surface brightness dimming.

W0831+0140, the highest redshift source, also shows an apparent gradient in the polarization fraction in Figure \ref{fg:new_HDs_2Dpol}, although the polarization angle is fairly constant across the source. However, the polarization fraction pattern is not clear when we separate by the different OBs (see Fig. \ref{fg:W0831_2D}), so it is possible that the pattern observed in Figure \ref{fg:new_HDs_2Dpol} is spurious. No clear polarization patterns are observed for W0220+0137 and W0116--0505 in any of the bands, suggesting that they are simply unresolved, which is consistent with their compactness as observed by \citet{assef20} in the {\it{HST}}/WFC3 F555W and F160W images. In the next sections we model the observed polarization fraction of these sources using different scattering media and geometries.  

\section{Discussion}\label{sec:discussion}

To explain the observed polarization fraction of W0116--0505 in the $R_{\rm Special}$ band, \citetalias{assef22} discussed two general scenarios for these objects. The first of these was a simple model in which the quasar shone on a diffuse, optically thin medium, likely the ISM of the host galaxy, through an opening in the toroidal dust distribution that is blocking our line-of-sight to the accretion disk. The second scenario assumed the scattering material was dust at the base of a massive ionized gas outflow powered by the quasar. \citetalias{assef25} probed the latter scenario further using radiative transfer simulations and found a good match to the W0204--0506 observations. Below we discuss the nature of the scattering material considering these models for our updated ensemble of observations. 

\subsection{Scattering by a Diffuse, Optically Thin Medium}\label{ssec:ism_scatt}

In this first scenario considered, the scattering properties will depend on the composition of the optically thin medium. For a fully ionized plasma, the process will be dominated by Thomson scattering off free electrons. For an ISM-like medium, the scattering would instead by dominated by dust grains. To analyze the latter quantitatively, we consider the dust mixtures studied by \citet{draine03} for the SMC bar, the LMC average, and the MW, which should span the range of plausible compositions for the host galaxy. While it is possible to have a combination of both free electrons and dust, the scattering optical depth of dust grains is much larger, so as long as dust is present, it will tend to dominate \citepalias[see, e.g., discussion in][]{assef22}.

\begin{figure*}
    \centering
    \includegraphics[width=\textwidth]{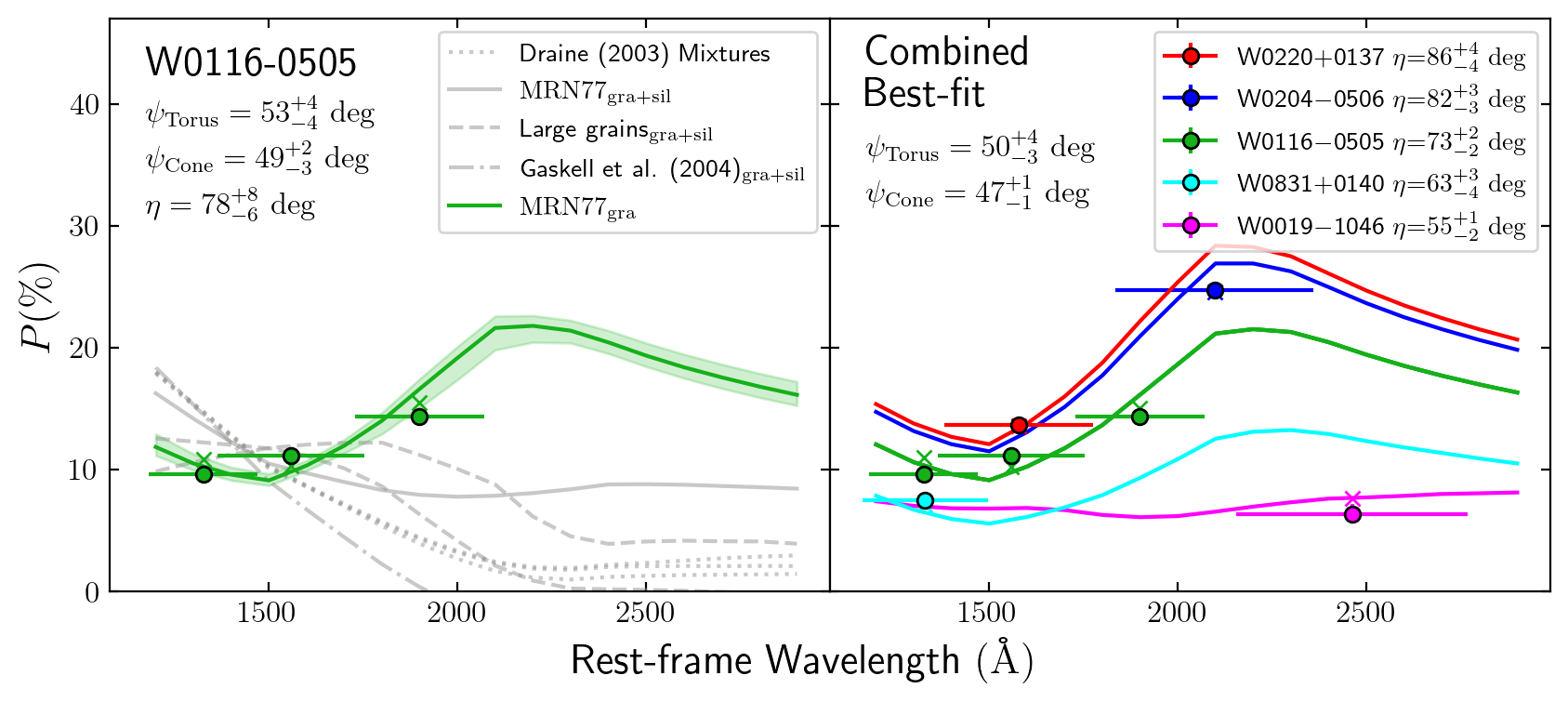}
    \caption{{\it{(Left panel)}} The solid circles show the observed broad-band polarization for the BHD W0116--0505 as a function wavelength. The lines show the predictions from the {\tt{SKIRT}} simulations for different dust mixtures assuming the scattering medium consists of dust in a polar outflow as discussed in \S\ref{ssec:scat_polar_outflow}. For each model we pick the predictions that best-fit the data by optimizing over the torus opening angle, $\psi_{\rm Torus}$, the polar outflow cone opening angle, $\psi_{\rm Cone}$ and the inclination angle of the system $\eta$. The only dust mixture that recovers the wavelength behavior of the observed polarization fractions corresponds to the graphite-only $\rm MRN77_{\rm gra}$ (see Table \ref{tab:SKIRT_params} for details). The best-fit model parameters for this dust mixture are shown in the top left corner, and the expected values measured in the broad-bands are shown by the crosses. {\it{(Right panel)}} The solid circles show the observed broad-band polarization for all BHDs studied in this article. The lines show the prediction from a polar outflow scatterer assuming the $\rm MRN77_{\rm gra}$ dust mixture, for the best-fit model that requires all BHDs to share torus and outflow polar cone opening angles. Only inclination is allowed to change between BHDs. The best-fit values for $\psi_{\rm Torus}$ and $\psi_{\rm Cone}$ are shown in the top left corner, while the best-fit inclination for each object is shown in the legend box. The crosses show the broad-band measurements expected from the model.}
    \label{fg:pfrac_wave}
\end{figure*}

Figure \ref{fg:pfrac_wave} shows the polarization fraction of each target in each observed broad-band as a function of rest-frame wavelength. We see that the polarization fraction rises towards longer wavelengths in W0116--0505. Since Thomson scattering is independent of wavelength, it is not possible to model the observed broad-band polarization fractions of W0116--0505 if we assume a simple scattering geometry where the accretion disk continuum and the broad-line region (BLR) emission have the same polarization properties. If their polarization properties were not the same, the different BLR and continuum contributions to each broad-band could result in different observed polarization fractions. We study further in Appendix \ref{app:thomson_thin_scatt} the case where they are scattered differently, but find we cannot reproduce the observed polarization fractions simultaneously with the lack of change in polarization angles between broad-bands.

Something similar happens with ISM dust scattering, as all dust mixtures considered predict the polarization fraction in W0116--0505 should be rising towards the blue, opposite to the observed trend (see Fig. \ref{fg:W0116_ISM_Dust}). We also find that the dust mixtures considered are unable to match the observed polarization fraction of $14.36\pm 0.43$ in the $I_{\rm Bessell}$ band, which is free of BLR contamination (see Fig. \ref{fg:SED_w0116}), for any inclination and torus opening angle. Furthermore, as already pointed out by \citetalias{assef25}, none of these mixtures is able to reproduce the polarization fraction of $24.72\pm 0.66$ observed in the $R_{\rm Special}$ band of W0204--0506. Further details are discussed in the Appendix \ref{app:dust_thin_scatt}.

Considering all this, we conclude it is unlikely that the scattering medium corresponds to optically thin ionized gas or ISM dust in the case of W0116--0505. Given the similarities between the SEDs of all the BHDs considered, it is unlikely that the mechanism responsible for the scattering is substantially different for the rest as compared to W0116--0505. In the next section we discuss the possibility that the scattering medium corresponds to a dusty quasar outflow which can have very different dust mixtures than the standard ISM of a galaxy, and we find a much better agreement with the observations.

\subsection{Scattering from a Dusty Polar Outflow}\label{ssec:scat_polar_outflow}

As discussed earlier, a significant fraction of Hot DOGs, including BHDs, are known to have powerful outflows detected using the [O{\sc iii}]$\lambda\lambda 5007, 4959$ emission lines \citep{wu18, jun20, finnerty20, vayner25} as well as some far-IR transitions \citep{fan18, martin24, liao24a}. Of the BHDs studied in this article, powerful ionized-gas outflows have been detected in W0116--0505, W0220+0137 and W0831+0140 by \citet{finnerty20} using the [O{\sc iii}] emission line. They estimated mass outflow rates of $\log{\dot{M}_{\rm Outflow}/M_{\odot}~\rm yr^{-1}} = 3.9\pm 0.3$ and $2.5\pm 0.3$ for W0116--0505 and W0220+0137 respectively. They did not estimate this quantity for W0831+0140 due to flux calibration issues despite a clear outflow component detected in the [O{\sc iii}] line with a FWHM of $6000\pm 1000~\rm km~\rm s^{-1}$ and a blue-shift of $-2800\pm 400~\rm km~\rm s^{-1}$. No such observations unfortunately exist to date for W0019--1046 and W0204--0506.

Considering the possibility that outflows are common within this population, \citetalias{assef22} proposed that the scattered AGN light detected in the $R_{\rm Special}$ band in W0116--0505 corresponded to light scattered by dust at the base of the very massive outflow detected by \citet{finnerty20}, and demonstrated the feasibility of this scenario with simple calculations. Here we expand on this idea following a similar approach to that presented by \citetalias{assef25} to study W0204--0506, assuming this type of geometry and considering a variety of dust mixtures. Specifically, we set up simulations using the code {\tt{SKIRT}}\footnote{\url{https://skirt.ugent.be}} \citep{camps15, camps20}, which carries out Monte Carlo radiative transfer calculations for a given geometry and composition of the scattering material and an illumination source. We take here the general aspects of the dust AGN setup used by \citet{stalevski16, stalevski17, stalevski23}. 

\begin{figure}
    \includegraphics[width=0.5\textwidth]{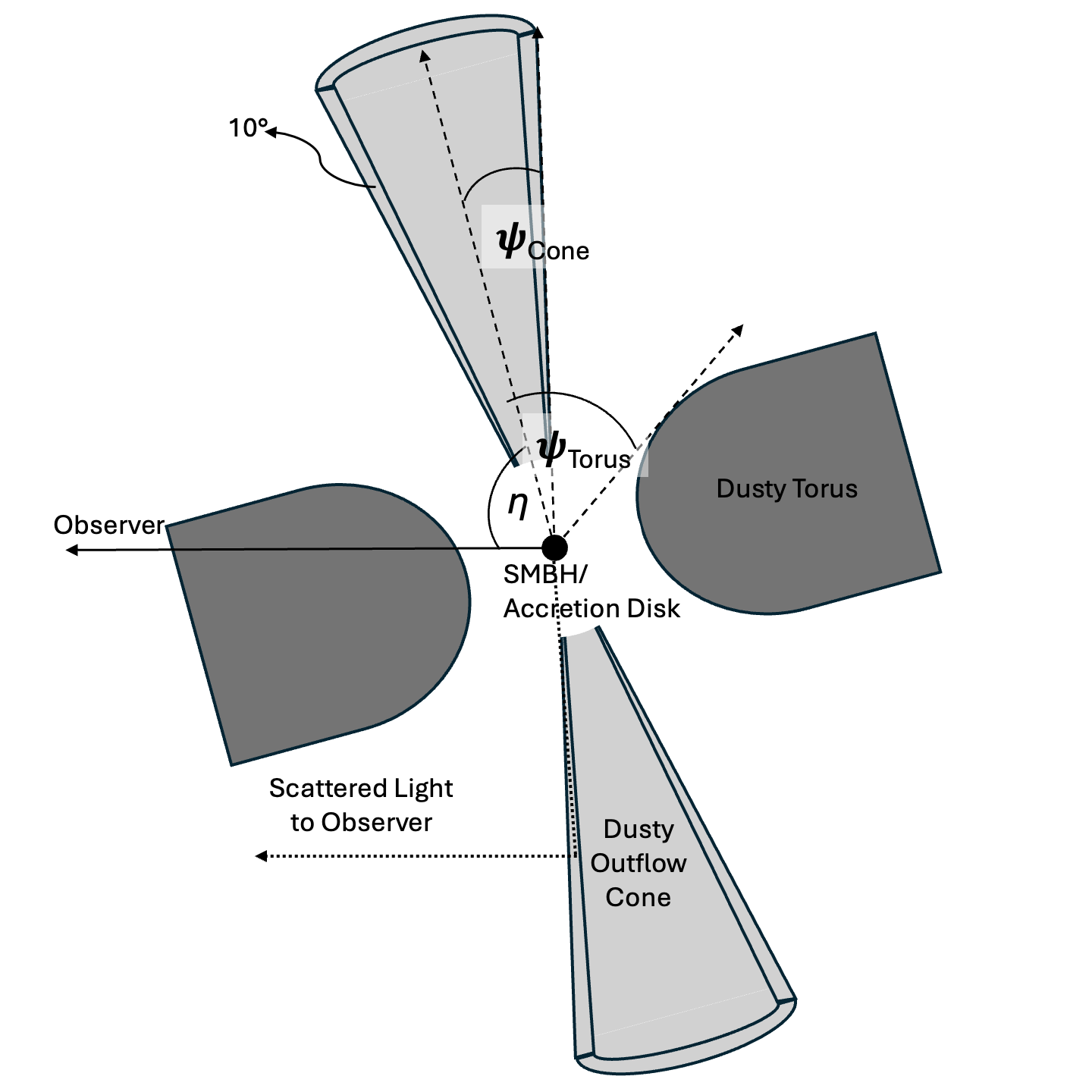}
    \caption{Cartoon of the geometry assumed for our {\tt{SKIRT}} simulations. Structures are not shown to scale.}
    \label{fg:cartoon}
\end{figure}

Our {\tt{SKIRT}} simulation consists of an accretion disk, assuming the SED of \citet{stalevski12}, surrounded by an optically thick dust torus with a half-opening angle $\psi_{\rm Torus}$, whose polar axis is inclined by an angle $\eta$ from the line-of-sight. Along the polar axis we add dust in a biconical shell distribution to emulate a biconical dust polar outflow. The cone has a half-opening angle of $\psi_{\rm Cone}$, with all the dust in the outermost ten degrees. Figure \ref{fg:cartoon} shows a cartoon of this geometry. As in \citetalias{assef25}, we assume a dust optical depth for a line of sight from the central engine through the dusty region of one of the cones to be $\tau=0.1$ at rest-frame 5500\AA. We find that the optical depth does not significantly affect the spectral shape of the scattered light, only strongly affecting the fraction of the central engine light that gets scattered. For $\tau=0.1$ we find a typical scattered light fraction is $\sim1\%$, consistent with what is typically observed for BHDs.

We use these simulations to first explore the polarization fraction produced by different dust mixtures and compare the predictions to the observations. As in the previous sections, we first focus on W0116--0505 since its multi-wavelength observations are ideal to differentiate between models. Given the large potential degeneracies between the parameters being fit \citepalias[see as well discussion in][]{assef25}, we first setup a fiducial geometry with a torus half-opening angle of $\psi_{\rm Torus}=50~\rm deg$ and a polar outflow cone half-opening angle of $\psi_{\rm Cone}=30~\rm deg$ and vary the inclination angles of the system between $\eta=55~\rm deg$ and 85~deg in steps of 10~deg. We run simulations for this range of inclinations and then interpolate between them to identify the value of $\eta$ that best reproduces the observations. In this initial broad search we explore the same array of dust mixtures explored by \citetalias{assef25}, which are detailed in Table \ref{tab:SKIRT_params}. Specifically, we start with a dust mixture that follows the size distribution of \citet[][hereafter \citetalias{mathis77}]{mathis77}, namely $\propto a^{-3.5}$, with  $a=0.005-0.25~\mu\rm m$ and has a silicon to graphite ratio of 51/49, as given by the normalization of the grain size distributions in \citet{weingartner01}. We refer to this mixture as $\rm MRN77_{\rm gra+sil}$. Additionally, we explore a mixture with the same grain size distribution and range but only considering graphites, which we refer to as $\rm MRN77_{\rm gra}$. We also consider two possibilities of larger dust grains with $a=0.1-1.0~\mu\rm m$ and $a=1.0-10.0~\mu\rm m$ but otherwise the same size distribution and silicate to graphite ratio of $\rm MRN77_{\rm gra+sil}$. For completeness we also consider the three dust mixtures of \citet{draine03} discussed in the previous section, as well as dust with the size distribution of \citet{gaskell04}.   

\begin{table*}
    \renewcommand{\arraystretch}{1.5}
    \caption{\label{tab:SKIRT_params} SKIRT Simulation Parameters}
    \small
    \begin{tabular}{lcp{0.7\textwidth}}
        \hline \hline
        \multicolumn{3}{l}{{\textbf{\textit{Grids for the Geometric Parameters}}}}\\
        Broad test of mixtures &:& $\psi_{\rm Cone} = 30~\rm deg$, $\psi_{\rm Torus} = 50~\rm deg$ and $\eta = 55-85~\rm deg$ in steps of 10~deg.\\
        Refined $\rm MRN77_{\rm gra}$ grid &:&  $\psi_{\rm Torus} = 25~\rm deg - 60~\rm deg$, $\psi_{\rm Cone} = 20~\rm deg - \psi_{\rm Torus}$ and $\eta = \psi_{\rm Torus} - 90~\rm deg$, all in steps of 5~deg.\\[5pt]

        \multicolumn{3}{l}{{\textbf{\textit{Dust Mixtures}}}}\\
        $\rm MRN77_{\rm gra+sil}$ &:&  Size distribution $\propto a^{-3.5}$, $a = 0.005 - 0.25~\mu\rm m$. Silicate to graphite ratio: 51/49.\\
        $\rm MRN77_{\rm gra}$ &:&  Same as $\rm MRN77_{\rm gra+sil}$ but only with graphites.\\
        Large grains$_{\rm gra+sil}$ &:& Same as $\rm MRN77_{\rm gra+sil}$ but with $a = 0.1 - 1.0~\mu\rm m$ and $a = 1.0 - 10.0~\mu\rm m$.\\
        Draine (2003) Mixtures &:& Dust and gas mixture models from \citet{draine03} for the SMC bar, the LMC average and the MW.\\
        Gaskell et al. (2004) $_{\rm gra+sil}$ &:& Dust mixture with the grain size distribution from \citet{gaskell04} and the same silicates to graphites ratio of $\rm MRN77_{\rm gra+sil}$.\\
        \hline
    \end{tabular}
\end{table*}

The left panel of Figure \ref{fg:pfrac_wave} shows the polarization fraction obtained for each dust mixture for the best-fit inclination to reproduce the observed polarization fraction of W0116--0505. The figure confirms the assessment from the previous section that the SMC-, LMC- and MW-type dust mixtures from \citet{draine03} cannot reproduce the observations. In addition to the difficulty in reproducing the observed $I_{\rm Bessel}$-band polarization, the observations cannot reproduce a polarization fraction that drops towards shorter wavelengths. The same is also true for the silicate-rich dust mixtures we tried, both with typical and large grain sizes as well as with the size distribution of \citet{gaskell04}. The graphite-only dust mixture $\rm MRN77_{\rm gra}$ does, however, a qualitatively good job of reproducing the observed trend. This is qualitatively consistent with the scattering medium consisting of a dusty polar outflow wind, as the high temperatures of the wind and its potential origin close to the sublimation zone would naturally result in graphite rich dust due to its higher sublimation temperature ($\sim$1800~K for graphites vs $\sim$1200~K for silicates).

Considering this, we now explore more in depth the geometry of the scattering medium assuming the \citetalias{mathis77} dust mixture and varying all three parameters, $\psi_{\rm Torus}$, $\psi_{\rm Cone}$ and $\eta$ to understand how well we can constrain them. Specifically, we compute models in a grid of $\psi_{\rm Torus}=25~\rm deg$ to 60~deg, $\psi_{\rm Cone}=20~\rm deg$ to $\psi_{\rm Torus}$, and $\eta=\psi_{\rm Torus}$ to 90~deg, all in steps of 5~deg. This is summarized in Table \ref{tab:SKIRT_params}. We note that we do not allow  $\eta<\psi_{\rm Torus}$ as the line of sight towards the accretion disk would not be blocked by the torus, resulting in a much lower polarization fraction due to dilution from the much brighter non-scattered light of the AGN. We also do not consider larger opening angles for the torus as they would be inconsistent with the large covering fractions required to explain their bright, obscured IR SEDs \citep{tsai15, tsai24}. We optimize the parameters through a Markov Chain Monte Carlo (MCMC) process using the {\tt{emcee}} algorithm \citep{emcee}, and adopt as the best-fit parameters the median of their distribution resulting from the MCMC. The left panel of Figure \ref{fg:pfrac_wave} shows the polarization fraction as a function of wavelength for the model obtained from the best-fit parameters, namely $\psi_{\rm Torus} = 53\pm 4~\rm deg$, $\psi_{\rm Cone} = 49^{+2}_{-3}~\rm deg$ and $\eta = 78^{+8}_{-6}~\rm deg$. The similar values of $\psi_{\rm Torus}$ and $\psi_{\rm Cone}$ imply a preferred model where there is little discontinuity between the dust from the outflow and the dust from the torus. The only unobscured sight-lines to the accretion disk would be those going through the outflow cone. This geometry is also qualitatively consistent with that of the dust torus model proposed by \citet{honig19}, in which a polar wind lifts dust from the sublimation region. 

While this model is able to qualitatively reproduce the observed polarization fractions well, we find that nominally the goodness-of-fit is high, with $\chi^2=36$. Searching for the combination of parameters that minimizes the $\chi^2$ statistics, we find a marginally better solution with $\chi^2_{\rm min}=29$. The issue seems to be that the models have a significant upturn at the shortest wavelengths mapped by the $v_{\rm High}$-band that is inconsistent with the observed broad-band polarization fractions. This may be caused by the significant contribution from Ly$\alpha$ to the $v_{\rm High}$ band (see Fig. \ref{fg:SED_w0116}), which accounts for about 30\% of the flux in the band. Unlike the other broad emission lines in the spectrum, Ly$\alpha$ may not be coming from the broad-line region, which is within the torus, but rather from the circum-galactic medium (CGM) through a number of different processes that would make this emission line unpolarized. Emission of Ly$\alpha$ at CGM scales has been observed for Hot DOGs \citep{bridge13, ginolfi22, shobhana26} as well as for other luminous quasars \citep[e.g.,][]{battaia19}. If we assume that all the Ly$\alpha$ flux observed in the spectrum is unpolarized, the intrinsic continuum polarization fraction in the $v_{\rm High}$ band would increase to $13.6\pm 0.5~\%$ and can be nearly perfectly fit by the model ($\chi^2_{\rm min} = 0.6$). Similarly, if we assume that all the broad the emission lines are polarized at the 5\% level as in the case of the ERQ SDSS1652+1728 \citep{alexandroff18} maintaining the same polarization angle as in the continuum, the intrinsic continuum polarization fraction in the $v_{\rm High}$ would increase to $16.1\pm 0.9~\%$ and be consistent with the models ($\chi^2_{\rm min} = 0.08)$. We find that while the fits improve in both cases, the qualitative results are not very sensitive to these corrections. Furthermore, it is important to note that the increased polarization fractions in the models towards shorter wavelengths at $\lambda < 1500\AA$ are likely due to scattering in the inner edge of torus and becomes very dependent on the exact amount of dust along the line of sight, which could be quite large in reality at torus grazing angles due to diffuse dust in the torus outskirts as well as dust concentrated along the base of the outflow cone (our SKIRT model assumes the dust is uniformly distributed). Hence we do not consider any corrections to the $v_{\rm High}$ band, but note that such details could potentially be answered through spectropolarimetry in the future. 

\begin{figure*}
    \centering
    \includegraphics[width=\textwidth]{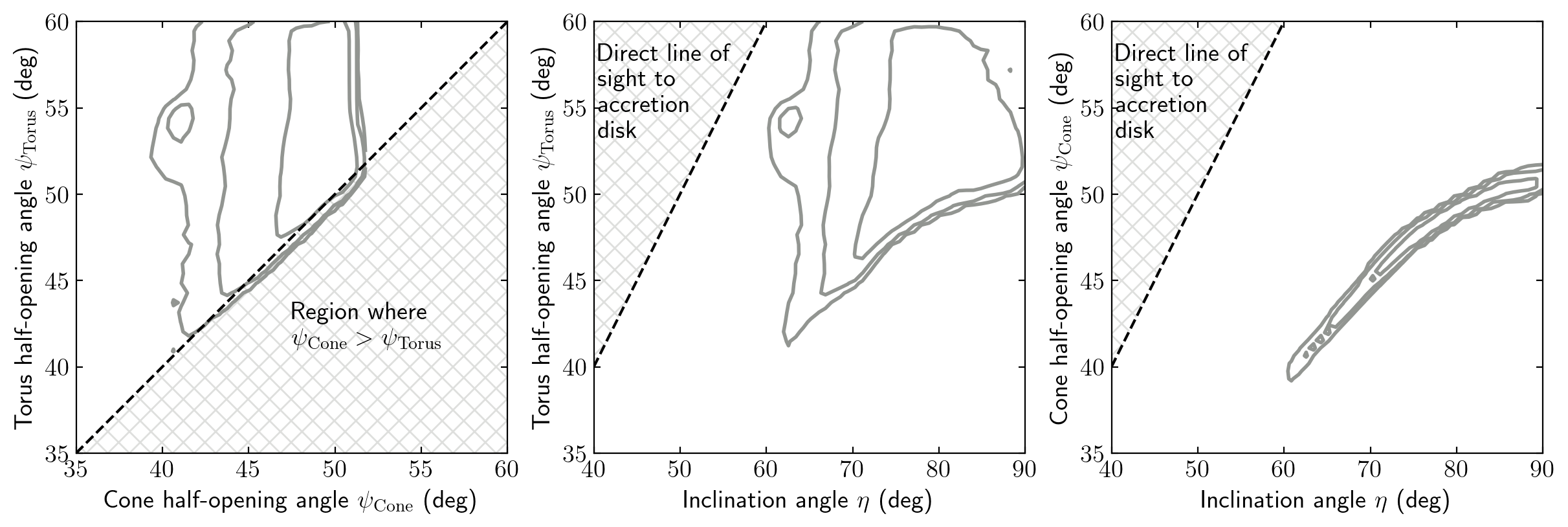}
    \caption{Probability contours (68.3\%, 95.4\% and 99.1\%) for the distribution of the best-fit parameters from our SKIRT models to the multi-wavelength photometry of W0116--0505. The shaded areas mark combinations of parameters that would not be possible, either because they would allow a direct line of sight to the accretion disk, or because they would require an outflow cone opening angle larger than that of the torus.}
    \label{fg:w0116_skirt_parcorr}
\end{figure*}

Note that while there is a wide range of values for all the parameters, they are heavily correlated, as shown in Figure \ref{fg:w0116_skirt_parcorr}. We see that the preferred models have outflow cone half-opening angles $\psi_{\rm Cone}$ that range from being equal to the torus half-opening angle $\psi_{\rm Torus}$ to being $\sim 10-15~\rm deg$ smaller. This suggests that there are limited lines of sight that would not go directly through either the torus or the outflow. If, in reality, the dust is mostly concentrated at the base of the outflow rather than being uniformly distributed in the cone shell like in our model, this may imply most lines of sight to the central engine are still obscured to the observer. This would be consistent with the conclusions drawn by previous studies \citep[e.g.,][also see \S\ref{sec:intro}]{tsai15,assef15,jun21,li24} that the obscuration in BHDs, and in Hot DOGs in general, is not simply due to an orientation effect like it would be in the unified AGN model \citep[e.g.,][]{antonucci93}, but rather probes a very different stage of SMBH accretion as compared to regular quasars. Dusty outflows have also been observed in other obscured quasars \citep[e.g.,][]{temple19, fawcett23, vayner24}, further supporting this scenario. The Figure also shows that there is a tight correlation between $\psi_{\rm Cone}$ and $\eta$, which just reflects the fact that the polarization fraction strongly increases for higher inclination angles and for smaller outflow cones, so matching the observed values requires a trade-off between them. 

Considering then that BHDs may correspond to the stage where the quasar-driven winds are starting to clear dust around the central engine, it makes sense to test whether there is a uniform set of parameters that could explain the polarization fractions observed for all targets studied here. If BHDs comprise a fairly uniform set of objects, then we would expect the inclination angle $\eta$ to be the only parameter to vary significantly between them. Hence, starting from the assumption that \citetalias{mathis77} dust is present in all targets, we fit for the torus and polar outflow cone half-opening angles that best match the ensemble of BHDs studied here, and find somewhat similar parameters as for W0116--0505 with best-fit values of $\psi_{\rm Torus} = 50^{+4}_{-3}~\rm deg$ and $\psi_{\rm Cone} = 47\pm 1~\rm deg$ with inclination angles $\eta$ ranging between 55 and 86~deg. The results are shown in the right panel of Figure \ref{fg:pfrac_wave}. The agreement is quite remarkable given the simplicity of the model, indeed suggesting that BHDs may comprise a rather uniform class of objects.

\begin{figure}
    \centering
    \includegraphics[width=0.49\textwidth]{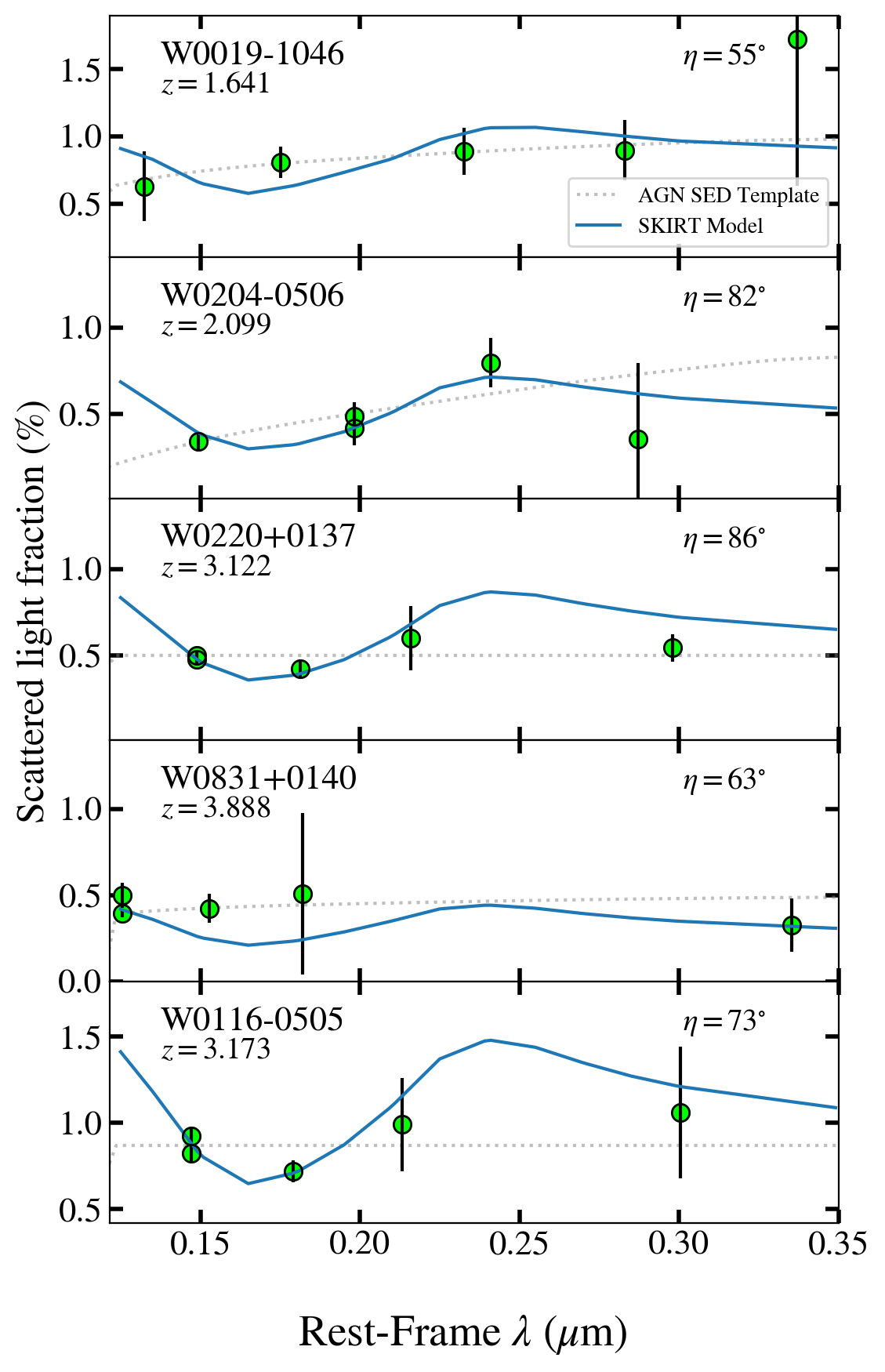}
    \caption{Scattered light fraction estimated as discussed in the text for each target in each rest-frame UV photometric band used to model their SEDs (green circles, see \S\ref{ssec:phot} for details). The dotted line shows the scattered light implied by the best-fit lightly reddened AGN continuum used in the SED modeling, while the blue solid line shows the scattered fraction estimated from the polar outflow SKIRT simulations discussed \S\ref{ssec:scat_polar_outflow}, with an amplitude estimated to best-fit the estimated scattered light fraction in the photometric bands.}
    \label{fg:scattered_light_fractions}
\end{figure}

Figure \ref{fg:scattered_light_fractions} shows the scattered-light spectrum expected for the best-fit inclination angle of each target using the best-fit {\tt{SKIRT}} model to the ensemble of BHDs as well as an estimate of the scattered light fraction as a function of wavelength for each target. The latter was created by, first, subtracting the best-fit host galaxy contribution from the rest-frame UV photometry to only leave the scattered light contribution. We then divide the resulting flux in each band by the flux expected for the hyper-luminous quasar powering the mid-IR if it was unobscured, assuming the \citet{assef10} AGN template. Notably, all the SKIRT predictions show a dip at $\lambda\sim1800~\rm\AA$. This is not to be confused with the graphite absorption feature at $2175~\rm\AA$ \citep{draine03a}, but rather the dip appears due to the enhanced scattering at slightly longer wavelengths, as dictated by the graphite optical properties. The equivalent width of this feature seems to be nearly independent of the geometry of the system within the range of the parameters probed. The presence of this feature seems to be qualitatively consistent with the scattered light SED shape of W0116--0505, W0204--0506 and W0220+0137, but it seems a featureless scattered light component may provide a better description of the observations in W0019--1046 and W0831+0140. This could be due to a significant contribution from other scattering mechanisms, like Thompson scattering from ionized gas, in the latter two objects, but could also just be a reflection of the discreteness and SNR of the photometry in addition to the coarseness of the model used to derive the scatter-light SEDs. More detailed multi-wavelength observations or high-SNR spectroscopy may further constrain the presence of this feature.

\section{Conclusions}\label{sec:conclusions}

In this article we discuss the polarization properties of a sample of BHDs. Specifically, we have obtained imaging polarization observations using the FORS2 instrument at the VLT in the $R_{\rm Special}$ band for the BHDs W0019--1046, W0220+0137 and W0831+0140, as well as in the $I_{\rm Bessel}$ and $v_{\rm High}$ bands for the BHD W0116--0505. We have added to this sample the $R_{\rm Special}$ imaging polarization observations of the BHDs W0116--0505 and W0204--0506, previously presented by \citetalias{assef22} and \citetalias{assef25} respectively. A summary of the target properties and observations is presented in Tables \ref{tab:targs} and \ref{tab:data}. 

We first measure the spatially integrated polarization properties of these targets, and find all of them to show significant polarization fractions, ranging between $P = 6.35\pm 0.79\%$ for W0019--1046 and $P = 24.72\pm 0.66\%$ for W0204--0506, both in the $R_{\rm Special}$ band (see Table \ref{tab:pols}). We also find that the polarization fraction of W0116--0505 rises from the $v_{\rm High}$ band ($P = 9.62\pm 0.38\%$) to the $I_{\rm Bessel}$ band ($P = 14.36\pm 0.43\%$), contrary to naive expectations for Thomson scattering off ionized gas, which should be wavelength independent, and for scattering off diffuse ISM dust, which should rise toward bluer wavelengths. We also analyze the spatial distribution of the polarization and find that W0019--1046 shows smooth gradients in the polarization angle and fraction in the $R_{\rm Special}$ band, with the latter rising from $\sim 3\%$ to $\sim 10\%$ toward the south-west, reminiscent to the gradient found by \citetalias{assef25} in W0204--0506 (see Figure \ref{fg:new_HDs_2Dpol}). No other robust gradients are identified.

We show that it is not possible to explain the wavelength dependence of the W0116--0505 polarization fraction through Thomson scattering, even when considering the potential contribution from the broad emission lines within the $R_{\rm Special}$ and $v_{\rm High}$ bands. This implies that the scattering medium must be dusty. We also show, however, that the observations cannot be explained through scattering off ISM-like dust. Using a simple model of an AGN shining into an smooth ISM through the opening angle of the torus, we find that a) the ISM dust scattering cannot reproduce the wavelength dependence of the W0116--0505, and that b) it is not possible to produce polarization fractions high enough to match the observed values in the $I_{\rm Bessel}$ band for W0116--0505 and in the $R_{\rm Special}$ band for W0204--0506. 

Following the approach presented by \citetalias{assef25} we also model the polarization fraction in the BHDs assuming that the scattering medium corresponds to dust in the outskirts of a polar outflow. The scattering in this model is simulated using the radiative transfer code {\tt{SKIRT}}. The assumed geometry is shown in Figure \ref{fg:cartoon}. We try a set of different dust mixtures and find that the only one that can qualitatively model the wavelength dependence of the polarization in W0116--0505 is one where the dust is dominated by graphites (see left panel of Fig. \ref{fg:pfrac_wave}), with best-fit geometric parameters of $\psi_{\rm Torus} = 53\pm 4~\rm deg$, $\psi_{\rm Cone} = 49^{+2}_{-3}~\rm deg$ and $\eta=78^{+8}_{-6}~\rm deg$. Interestingly, the preferred model has very similar values of $\psi_{\rm Torus}$ and $\psi_{\rm Cone}$, implying little discontinuity between the dust from the outflow and the dust from the torus, and that most lines of sight to the central engine will have some obscuration with the exception of those that go through the center of the outflow. Since in reality most of the dust may be concentrated at the base of the outflow, the effective covering fraction may be even larger than that suggested by the model. Additionally we find that we can simultaneously explain the current observations of all BHDs by assuming a single torus-cone geometry and only varying the inclination angle $\eta$ (see Fig. \ref{fg:pfrac_wave}). The radiative transfer simulations show that graphite-only dust would imprint a clear absorption feature in the scattered light at $\lambda\sim 1800$\AA, and we find indications that this feature is present in the broad-band UV observations of three out of the five BHDs studied.

Our analysis suggests that BHDs are a cohesive population of objects that share similar geometries, consistent with them probing a short-lived evolutionary state in the transition between a fully obscured state, when they appear as Hot DOGs, towards a normal quasar. This is consistent with the evolutionary sequences of obscured-to-unobscured quasars proposed by, for example, \citetalias{assef22}, \citet{li24} and \citet{stepney26b}. Our analysis also shows that polarimetry of hyper-luminous obscured quasars can be a powerful tool to explore the properties of outflows that is very complementary to the view obtained through traditional ionized gas probes such as broad [O{\sc iii}]. A combination of spatially-resolved, multi-wavelength polarization observations to fully trace the geometry of the outflows in combination with ionized gas to probe the line-of-sight kinematics can result in an exquisitely detailed view of quasar feedback. 

\begin{acknowledgements}
We thank Miika Pursiainen and Giorgos Leloudas for pointing out an issue in our previous work that had led to the overestimation of the uncertainties in the linear polarization fractions. RJA was supported by FONDECYT grant number 1231718. RJA, FEB, ML and AS acknowledge support by the ANID BASAL project FB210003. MS was supported by the Ministry of Science, Technological Development and Innovation of the Republic of Serbia (MSTDIRS) through contract no. 451-03-33/2026-03/200002 with the Astronomical Observatory (Belgrade). FEB acknowledges support from FONDECYT Regular 1241005. RFA is supported by the European Research Council (ERC) under the European Union’s Horizon 2020 research and innovation programme (DistantDust, Grant agreement No. 101117541). ML was supported by the Rubin-Chile Fund DIA3324. ES acknowledges support by the Alexander von Humboldt Foundation through a Humboldt Research Fellowship. The work of DS was carried out at the Jet Propulsion Laboratory, California Institute of Technology, under a contract with the National Aeronautics and Space Administration (80NM0018D0004). DJW acknowledges support from the Science and Technology Facilities Council (STFC; grant code ST/Y001060/1). This research made use of Photutils, an Astropy package for detection and photometry of astronomical sources \citep{photutils}.

\end{acknowledgements}

\bibliographystyle{aa}
\bibliography{refs.bib}

\begin{appendix}

\section{Details of the Observations}\label{app:obs}

A summary of the observations is presented in Table \ref{tab:data}.

\begin{table*}
    \caption{\label{tab:data} Imaging Polarimetry Observations}

    \begin{tabular}{lcccccccc}
        \hline \hline\\
        Target & 
        R.A. &
        Dec. &
        Program ID &
        Band &
        OB &
        Mean MJD & 
        Mean &
        Mean Seeing\tablefootmark{a}\\
        &
        (J2000) &
        (J2000) &
        &
        &
        &
        (days) &
        Airmass &
        (arcsec)\\\\
        \hline\\
        \multicolumn{5}{l}{\bf{BHDs}}\\
        W0019--1046 & 00:19:26.88 & --10:46:33.3 & 111.24UL.002 & $R_{\rm Special}$ & 3565057 & 60201.29 & 1.114 & 0.97\\
                    &             &              &              &                   & 3565577 & 60202.19 & 1.051 & 0.85\\
        W0116--0505 & 01:16:01.41 & --05:05:04.1 & 106.218J.001 & $R_{\rm Special}$ & 2886768 & 59135.18 & 1.064 & 0.57\\
                    &             &              &              &                   & 2886765 & 59136.17 & 1.069 & 0.84\\
                    &             &              &              &                   & 2886772 & 59136.20 & 1.073 & 0.85\\
                    &             &              &              &                   & 2886622 & 59137.17 & 1.066 & 0.91\\
                    &             &              & 111.24UL.001 & $v_{\rm High}$    & 3564847 & 60143.39 & 1.090 & 0.60\\
                    &             &              &              & $I_{\rm Bessel}$  & 3564862 & 60146.41 & 1.066 & 0.32\\
                    &             &              &              &                   & 3564862 & 60148.41 & 1.064 & 0.60\\
                    &             &              &              &                   & 3564862 & 60201.24 & 1.076 & 0.91\\
                    &             &              &              &                   & 3565005 & 60148.37 & 1.101 & 0.67\\
        W0204--0506 & 02:04:46.13 & --05:06:40.8 & 111.24UL.002 & $R_{\rm Special}$ & 3565580 & 60207.35 & 1.149 & 0.41\\
                    &             &              &              &                   & 3565504 & 60209.33 & 1.113 & 0.73\\
        W0220+0137  & 02:20:52.12 &  +01:37:11.6 & 111.24UL.002 & $R_{\rm Special}$ & 3565583 & 60201.33 & 1.129 & 0.89\\
                    &             &              &              &                   & 3565623 & 60207.30 & 1.120 & 0.43\\
        W0831+0140  & 08:31:53.25 &  +01:40:10.8 & 111.24UL.002 & $R_{\rm Special}$ & 3565626 & 60290.31 & 1.124 & 0.53\\
                    &             &              &              &                   & 3565639 & 60291.30 & 1.129 & 0.61\\
        \\
        \multicolumn{5}{l}{\bf{Polarization Standards}}\\\\
        BD--12 5133 & 18:40:01.60 & --12:24:07.4 & 60.A--9203(E)& $R_{\rm Special}$ &200277970& 59133.98 & 1.117 & 1.19\\
                    &             &              &              & $v_{\rm High}$    &         & 60148.00 & 1.565 & 0.64\\
        Vela1       & 09:05:59.90 & --47:18:57.9 &              & $I_{\rm Bessel}$  &200277985& 60238.34 & 1.514 & 1.05\\
                    &             &              &              & $R_{\rm Special}$ &         & 60238.34 & 1.537 & 0.95\\
        \\
        \multicolumn{5}{l}{\bf{Zero-Polarization Standards}}\\\\
        WD 0310--688& 03:10:31.30 & --68:36:05.3 &              & $R_{\rm Special}$ &200277988& 59117.36 & 1.410 & 1.58\\
        WD 1344+106 & 13:47:23.00 &  +10:21:34.8 &              & $v_{\rm High}$    &200277994& 60147.98 & 1.275 & 0.60\\
                    &             &              &              & $I_{\rm Bessel}$  &         & 60147.99 & 1.288 & 0.57\\
        WD 2039--202& 20:42:35.10 & --20:04:37.9 &              & $R_{\rm Special}$ &200278006& 59134.00 & 1.003 & 0.70\\
                    &             &              &              &                   &         & 60201.15 & 1.106 & 0.71\\
        WD 2359--434& 00:02:12.0  & --43:10:11.2 &              & $I_{\rm Bessel}$  &200278012& 60201.20 & 1.057 & 0.88\\
                    &             &              &              & $R_{\rm Special}$ &         & 60201.20 & 1.058 & 0.93\\
                    &             &              &              &                   &         & 60293.01 & 1.086 & 1.06\\
        \hline
    \end{tabular}
    \tablefoot{
        \tablefoottext{a}{Nominal value from seeing monitor obtained from the image headers.}
    }
\end{table*}

\section{Polarization Measurements}

A summary of the measurements obtained for each OB of each target, as well as for each of the standard stars is presented in Table \ref{tab:pols}.

\begin{table}
    \caption{\label{tab:pols} Polarization Measurements}
    \begin{small}
    \begin{tabular}{lclcc}
        \hline \hline\\
        Target & Band & OB &  P   & $\chi$ \\
               &      &    & (\%) &  (deg) \\
        \\
        \hline\\
        {\bf{BHDs}}\\
        W0019--1046  & $R_{\rm Special}$ & 3565057   & \phantom{0}7.14$\pm$1.09 &           166.6$\pm$4.5 \\
                     &                   & 3565577   & \phantom{0}5.31$\pm$0.80 &           161.7$\pm$4.7 \\
                     &                   & Combined  & \phantom{0}6.35$\pm$0.79 &           164.5$\pm$3.5 \smallskip\\
        W0116--0505  & $I_{\rm Bessel}$  & 3564862   &           13.93$\pm$0.78 & \phantom{0}73.6$\pm$1.7 \\
                     &                   & 3564862   &           14.64$\pm$0.73 & \phantom{0}73.4$\pm$1.4 \\
                     &                   & 3564862   &           14.19$\pm$0.85 & \phantom{0}71.2$\pm$1.7 \\
                     &                   & 3565005   &           14.39$\pm$0.70 & \phantom{0}72.4$\pm$1.4 \\
                     &                   & Combined  &           14.36$\pm$0.43 & \phantom{0}72.7$\pm$0.9  \smallskip\\
                     & $R_{\rm Special}$ & 2886622   &           10.87$\pm$0.52 & \phantom{0}75.2$\pm$1.4 \\
                     &                   & 2886765   &           11.50$\pm$0.43 & \phantom{0}75.1$\pm$1.1 \\
                     &                   & 2886768   &           11.87$\pm$0.48 & \phantom{0}72.3$\pm$1.1 \\
                     &                   & 2886772   &           10.55$\pm$0.42 & \phantom{0}73.8$\pm$1.2 \\
                     &                   & Combined  &           11.11$\pm$0.22 & \phantom{0}73.9$\pm$0.6  \smallskip\\
                     & $v_{\rm High}$    & 3564847   & \phantom{0}9.62$\pm$0.38 & \phantom{0}73.4$\pm$1.1  \smallskip\\
        W0204--0506  & $R_{\rm Special}$ & 3565504   &           25.20$\pm$0.96 & \phantom{0}12.4$\pm$1.1 \\
                     &                   & 3565580   &           24.30$\pm$0.81 & \phantom{0}13.0$\pm$1.0 \\
                     &                   & Combined  &           24.72$\pm$0.66 & \phantom{0}12.7$\pm$0.8 \smallskip\\
        W0220+0137   & $R_{\rm Special}$ & 3565583   &           13.70$\pm$0.54 &           150.6$\pm$1.1 \\
                     &                   & 3565623   &           13.60$\pm$0.45 &           150.2$\pm$0.9 \\
                     &                   & Combined  &           13.66$\pm$0.39 &           150.4$\pm$0.8 \smallskip\\
        W0831+0140   & $R_{\rm Special}$ & 3565626   & \phantom{0}7.67$\pm$0.56 &           131.9$\pm$2.1 \\
                     &                   & 3565639   & \phantom{0}7.21$\pm$0.51 &           131.2$\pm$2.1 \\
                     &                   & Combined  & \phantom{0}7.45$\pm$0.40 &           131.5$\pm$1.6 \\
        \\
        \multicolumn{5}{l}{\bf{Polarization Standards}}\\\\
        BD--12 5133  & $R_{\rm Special}$ & 200277970 & \phantom{0}4.13$\pm$0.03 &           146.5$\pm$0.2 \\
                     & $v_{\rm High}$    & 200277970 & \phantom{0}4.17$\pm$0.06 &           146.6$\pm$0.4 \smallskip\\
        Vela1        & $I_{\rm Bessel}$  & 200277985 & \phantom{0}7.13$\pm$0.03 &           171.6$\pm$0.1 \\
                     & $R_{\rm Special}$ & 200277985 & \phantom{0}7.79$\pm$0.06 &           171.8$\pm$0.2 \\
        \\
        \multicolumn{5}{l}{\bf{Zero-Polarization Standards}}\\\\
        WD 0310--688 & $R_{\rm Special}$ & 200277988 & \phantom{0}0.05$\pm$0.03 &          134.1$\pm$22.7 \smallskip\\
        WD 1344+106  & $I_{\rm Bessel}$  & 200277994 & \phantom{0}0.16$\pm$0.10 & \phantom{0}99.7$\pm$70.9 \\
                     & $v_{\rm High}$    & 200277994 & \phantom{0}0.23$\pm$0.12 &          178.3$\pm$15.6 \smallskip\\
        WD 2039--202 & $R_{\rm Special}$ & 200278006 & \phantom{0}0.14$\pm$0.07 &          116.2$\pm$15.7 \\
                     &                   & 200278006 & \phantom{0}0.04$\pm$0.03 & \phantom{0}95.5$\pm$66.6 \smallskip\\
        WD 2359--434 & $I_{\rm Bessel}$  & 200278012 & \phantom{0}0.23$\pm$0.12 & \phantom{0}20.1$\pm$18.3 \\
                     & $R_{\rm Special}$ & 200278012 & \phantom{0}0.14$\pm$0.10 & \phantom{0}28.0$\pm$29.0 \\
                     &                   & 200278012 & \phantom{0}0.33$\pm$0.15 & \phantom{0}10.7$\pm$15.0 \\
        \hline
    \end{tabular}
    \end{small}
\end{table}        

\section{All Spatially Resolved Observations}\label{app:2d_pol}

\begin{figure}[!h]
    \centering
    \includegraphics[width=0.49\textwidth]{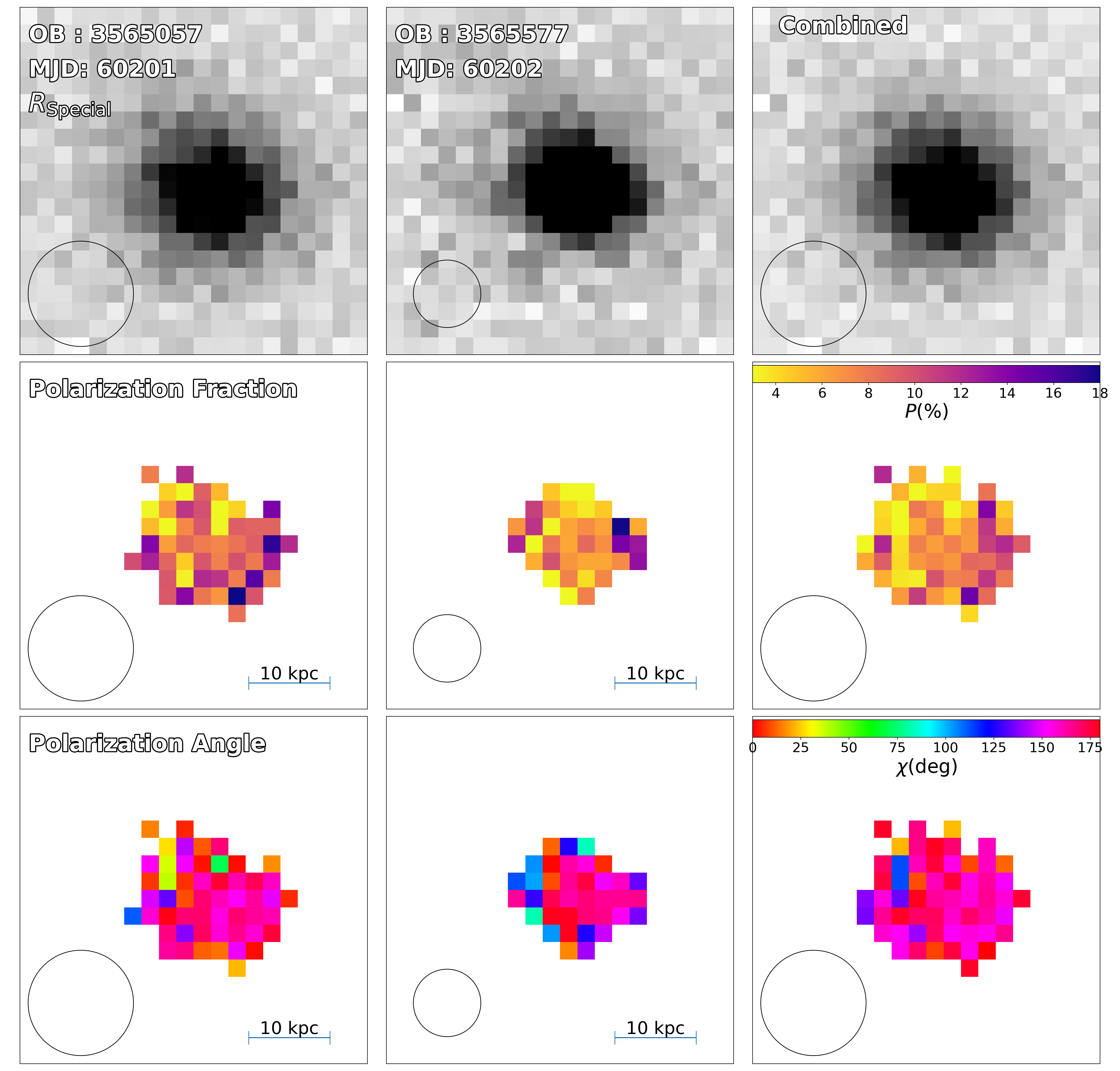}
    \caption{Polarization map of W0019--1046.}
    \label{fg:W0019_2D}
\end{figure}

\begin{figure}[!h]
    \centering
    \includegraphics[width=0.49\textwidth]{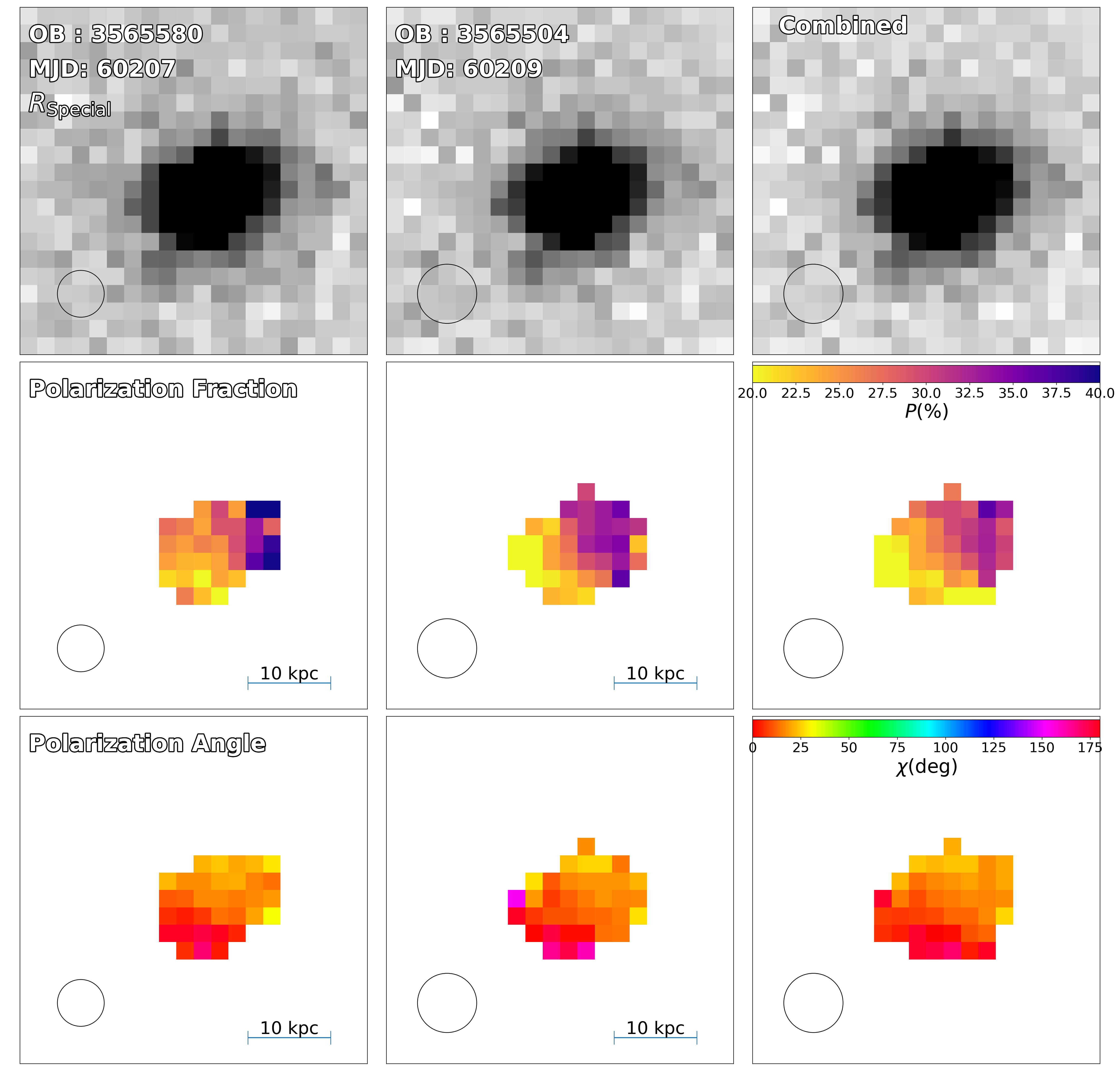}
    \caption{Polarization map of W0204--0506.}
    \label{fg:W0204_2D}
\end{figure}

\begin{figure}[!h]
    \centering
    \includegraphics[width=0.49\textwidth]{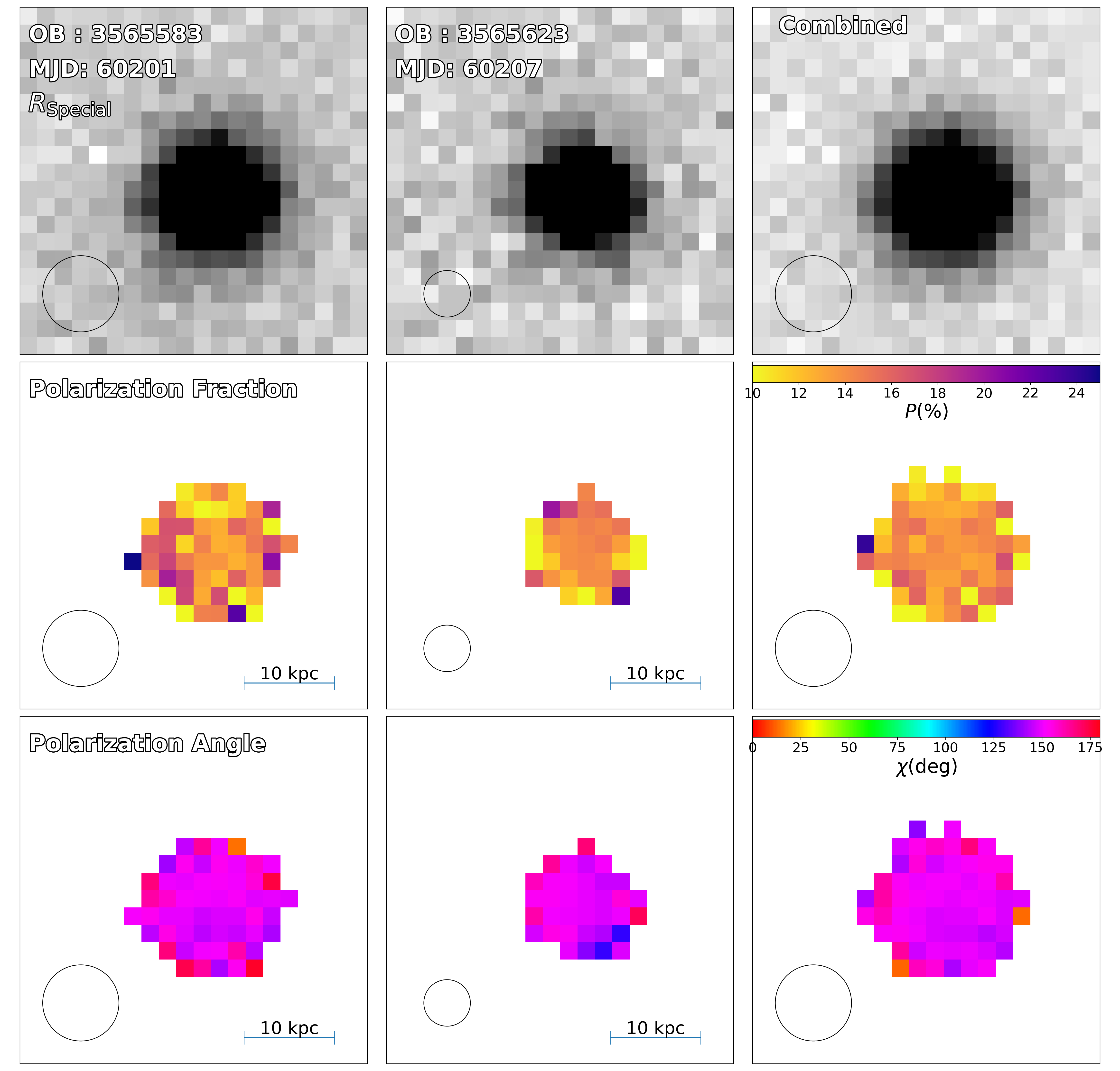}
    \caption{Polarization map of W0220+0137.}
    \label{fg:W0220_2D}
\end{figure}

\begin{figure}[!h]
    \centering
    \includegraphics[width=0.49\textwidth]{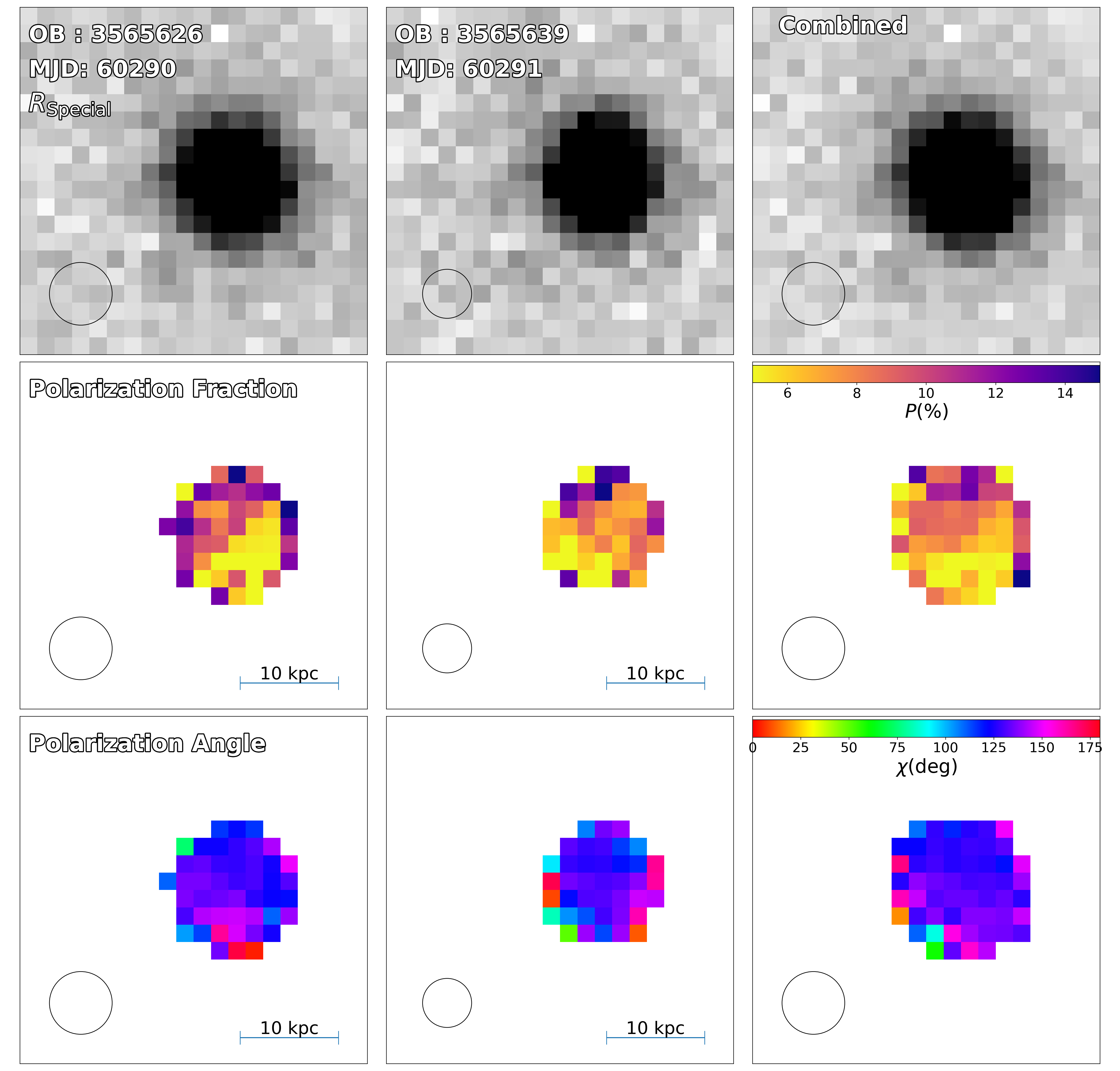}
    \caption{Polarization map of W0831+0140.}
    \label{fg:W0831_2D}
\end{figure}

\begin{figure}[!h]
    \centering
    \includegraphics[width=0.49\textwidth]{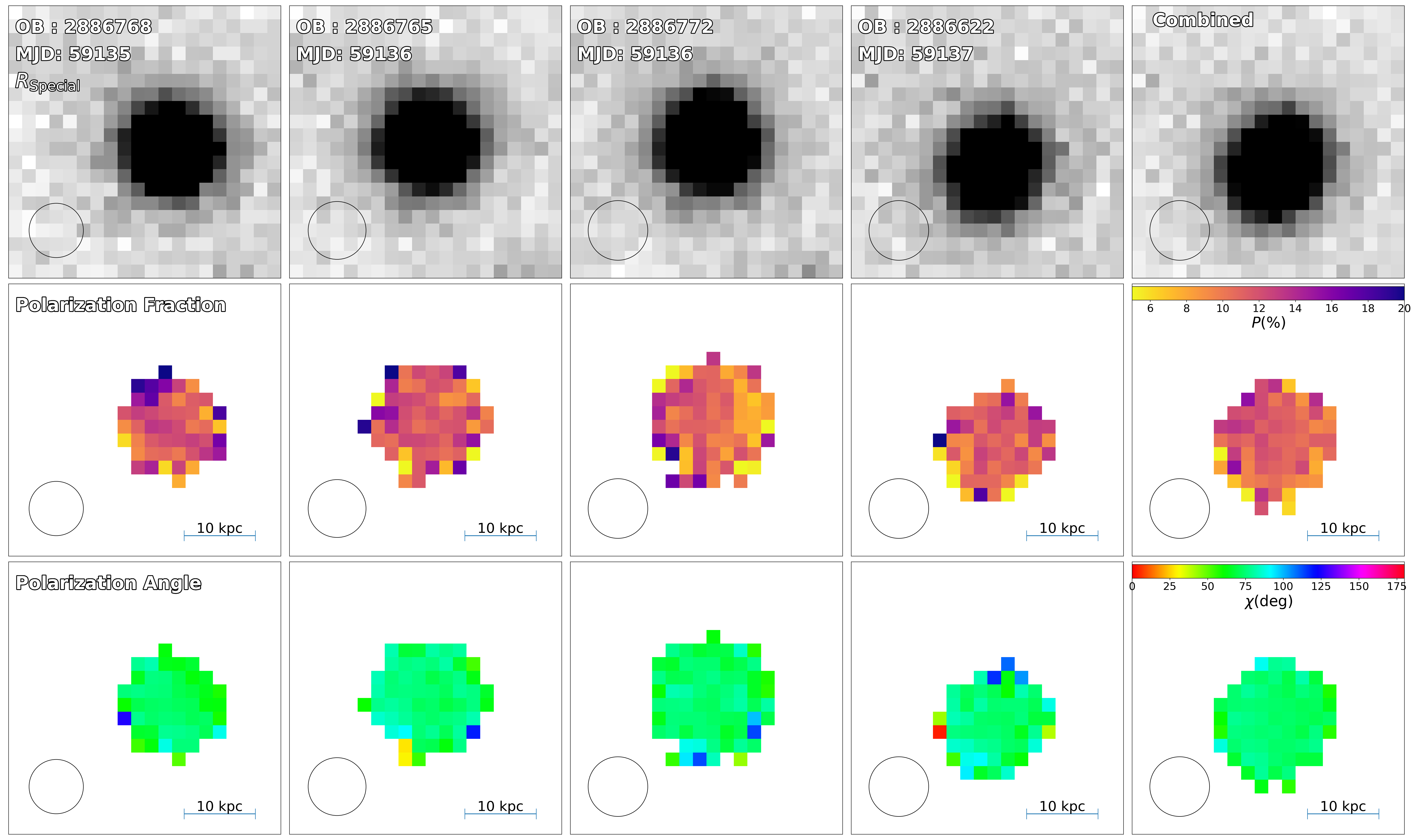}
    \caption{Polarization map of W0116--0505 in the $R_{\rm Special}$ band.}
    \label{fg:W0116_R_2D}
\end{figure}

\begin{figure}[!h]
    \centering
    \includegraphics[width=0.49\textwidth]{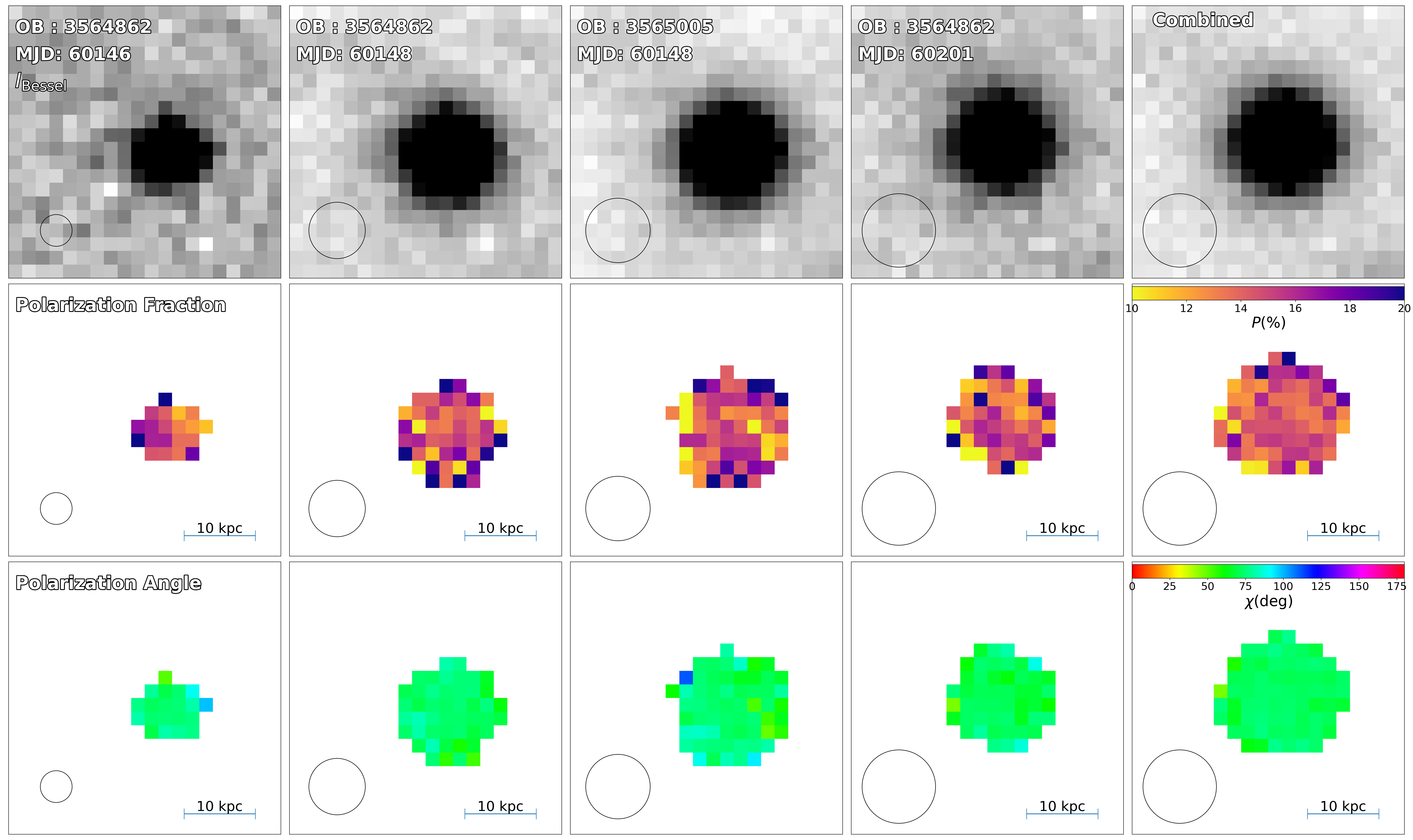}
    \caption{Polarization map of W0116--0505 in the $I_{\rm Bessel}$ band.}
    \label{fg:W0116_I_2D}
\end{figure}

\begin{figure}[!h]
    \centering
    \includegraphics[width=0.4\textwidth]{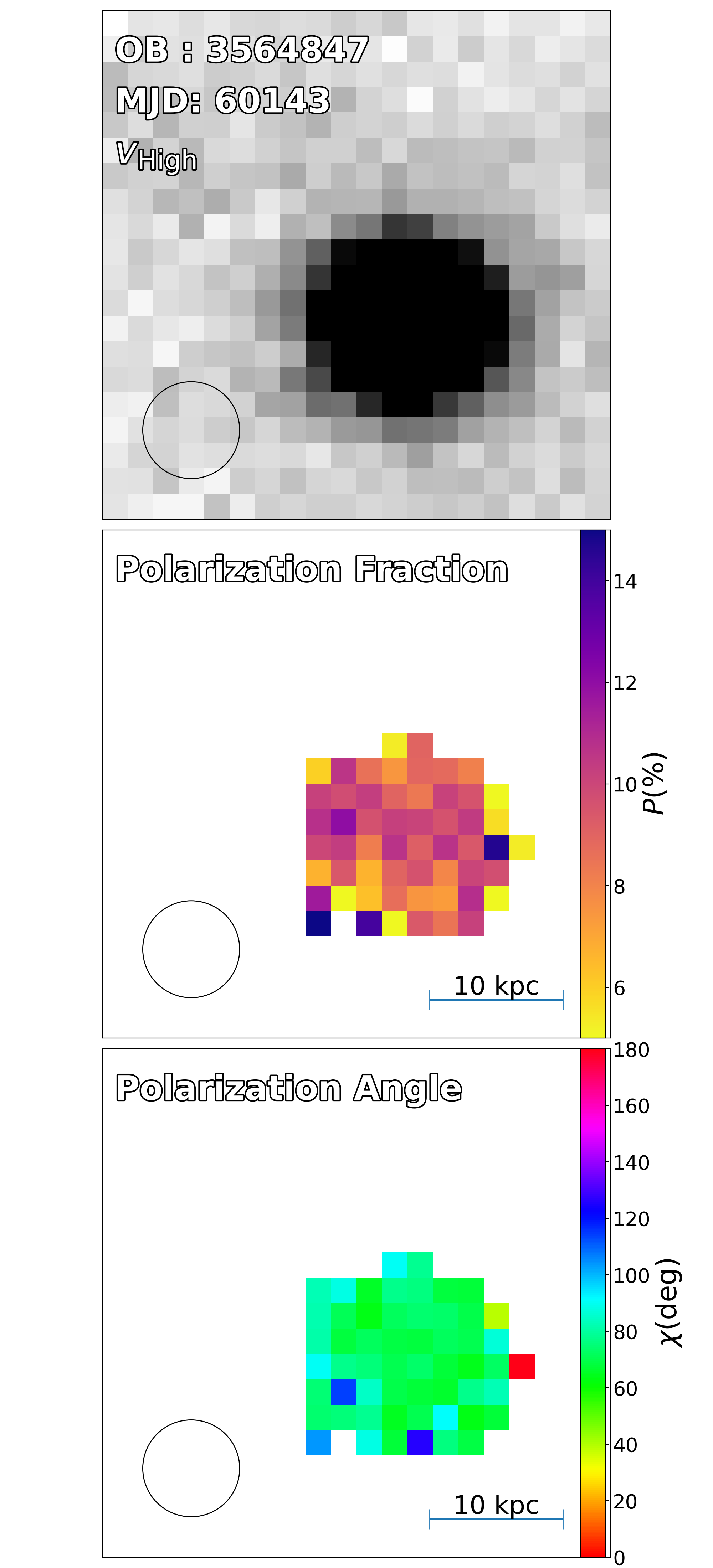}
    \caption{Polarization map of W0116--0505 in the $v_{\rm High}$ band.}
    \label{fg:W0116_V_2D}
\end{figure}

Here we present polarizations maps for all the imaging polarization observations discussed in the main text. We show the polarization fraction and angle spatial maps for each individual OB and for the combined observations for the $R_{\rm Special}$ observations of W0019--1046 (Fig. \ref{fg:W0019_2D}), W0204--0506 (Fig. \ref{fg:W0204_2D}), W0220+0137 (Fig. \ref{fg:W0220_2D}) and W0831+0140 (Fig. \ref{fg:W0831_2D}). We also show the maps for the $R_{\rm Special}$ (Fig. \ref{fg:W0116_R_2D}) $I_{\rm Bessel}$ (Fig. \ref{fg:W0116_I_2D}) and $v_{\rm High}$ (Fig. \ref{fg:W0116_V_2D}) observations of W0116--0505. 

\section{Gas Scattering}\label{app:thomson_thin_scatt}

We investigate here whether the observed polarization fractions can be explained through Thomson scattering from free electrons in an ionized gas medium. As Thomson scattering can reproduce any polarization fraction at a single wavelength, few constraints can be obtained by looking at monochromatic observations of a single target. Hence, we focus here first on W0116--0505, for which polarization fractions are measured in three different broad-bands and can break some of these degeneracies.  

As discussed in Section \ref{ssec:ism_scatt}, we see that the polarization fraction rises towards longer wavelengths in W0116--0505. Since Thomson scattering is independent of wavelength, it is not possible to model the observed broad-band polarization fractions of W0116--0505 if we assume a simple scattering geometry, where all components from the AGN are scattered with the same polarization fraction and angles. 

In principle, it is possible that the broad emission lines are resonantly scattered by the gas medium. This could lead to a different scattering geometry, and hence different polarization fractions or angles with respect to the continuum, as highlighted by \citet{zakamska23}, particularly in the case where the line-emitting region is outflowing. The difference in polarization fractions between the different broad-bands could be a result of the combination of the continuum and broad emission lines in each band. We consider this possibility through a simple test fitting separately for the polarization fraction and angles of the emission lines and the continuum. We separate the components by modeling the spectrum of W0116--0505 in the same manner as discussed in \citetalias{assef22}. Namely, we fit the continuum with a power-law and the observed emission lines using Gaussian functions. The C{\sc iv}, N{\sc v} and the Si{\sc iv}-O{\sc iv}] blended emission are well modeled assuming a single Gaussian function, but we need to use two Gaussian functions to properly model the Ly$\alpha$ emission, as found by \citet{diaz-santos21} for this target. This model is shown in Figure \ref{fg:SED_w0116}. Note that the $I_{\rm Bessel}$ band does not overlap with any strong emission lines and hence provides, in this model, an estimate of the continuum polarization.

We assume all emission lines have the same polarization fraction and angle, $p_l$ and $\chi_l$, which can differ from those of the continuum, $p_c$ and $\chi_c$, and we find these need to be substantially different. The best-fit model requires the polarization angles to differ by 60$\pm$3~deg and that the emission lines have a polarization fraction of 42$\pm$4\%, about 3 times the 14.3$\pm$0.4\% polarization fraction of the continuum. The reason this is a natural solution is because the contribution from the emission lines with respect to the continuum is higher in the $v_{\rm High}$ than in the $R_{\rm Special}$ band, and the different polarization angles make the emission lines and continuum polarization combine destructively, lowering the integrated broad-band polarization fractions and allowing the model to reproduce the tendency observed in Figure \ref{fg:pfrac_wave}. We can discard this possibility, however, as this would require the polarization angles in $v_{\rm High}$ and $R_{\rm Special}$ to differ by 42$\pm$4~deg and 33$\pm$4~deg with respect to the one measured in $I_{\rm Bessel}$. Yet, as shown in Table \ref{tab:pols_short}, all three angles are measured to be consistent within their uncertainties.

\citet{alexandroff18} shows that the polarization signal in the ERQ SDSS1652+1728 can be approximately described by a continuum polarization fraction of about 15\% (with a small trend towards larger polarization at shorter wavelengths) and a broad-line polarization fraction of about 5\%, with little difference between their polarization angles. If we try to fit the observed broad-band values in W0116--0505 by allowing for an intrinsically different polarization fraction between the lines and the continuum but forcing the polarization angle to be the same, we find that we cannot satisfactorily reproduce the observations. While the best-fit polarization fractions are 13.8$\pm$0.4\% for the continuum and 6$\pm$1\% for the emission lines are reminiscent of the case for the ERQ SDSS1652+1728, the model has $\chi^2=33$. 

In addition, we note that \citetalias{assef22} showed that the {\it{HST}}/WFC3 F555W observations of W0116--0505 are resolved, and are extended in a direction that is nearly perpendicular to the polarization angle, suggesting that the polarization is coming from scattering on kpc scales where it is very unlikely to be dominated by free-electrons from ionized gas instead of dust. 

Considering all this, we conclude that ionized gas is unlikely to be the dominant scattering medium in the case of W0116--0505. Given the similarities between the SEDs of all the BHDs considered and that the polarization map is extended on kpc-scales in W0019--1046 and W0204--0506 (see \S\ref{ssec:2d_pol} and \citetalias{assef25} for details), it is unlikely that the mechanism responsible for the scattering is substantially different for the rest as compared to W0116--0505. Hence, we conclude that scattering from electrons in a fully ionized gas medium is an unlikely explanation for the polarization of BHDs in general and do not consider this scenario further. 

\section{ISM Dust Scattering}\label{app:dust_thin_scatt}

To explain the $R_{\rm Special}$ polarization fraction of W0116--0505, \citetalias{assef22} considered a simple model in which an AGN, consisting of an accretion disk surrounded by a toroidal dust structure with a given opening half-angle ($\psi_{\rm Torus}$) and inclination angle ($\eta$), shines onto an spherically symmetric ISM that is optically thin (i.e., photons suffer at most one scattering event). \citetalias{assef22} tested the possibility of this ISM consisting of the three dust mixtures studied by \citet{draine03} which are representative of the dust in the Milky Way (MW), LMC and SMC.

Given the strong wavelength dependence of the dust polarization fractions and scattering cross-sections of the dust mixtures provided by \citet{draine03}\footnote{We note that the functions for the polarization fraction and scattering cross-section provided by \citet{draine03} become somewhat uncertain at $\lambda<2700$~\AA\ due to the approximation used for the phase function. Full radiative transfer calculations are required to avoid this uncertainty, and we discuss those in section \ref{ssec:scat_polar_outflow}, yet we find the results are not qualitatively different.}, we slightly modify the equations of \citetalias{assef22} for this model. Instead of looking at the monochromatic polarization at the effective wavelength of the band, we calculate the convolution of the Stokes parameters as a function of wavelength with the respective filter curves. This provides a more reliable comparison to the measured broad-band values, although we note that the differences are modest. Furthermore, motivated by the asymmetric distribution of the polarization in W0204--0506 \citepalias{assef25}, we also consider models where we only see forward or backward scattering. For each object, dust mixture and direction of scattering we find the values of $\psi_{\rm Torus}$ and $\eta$ that minimize the $\chi^2$ with the observed data.  

\begin{figure}
    \centering
    \includegraphics[width=0.49\textwidth]{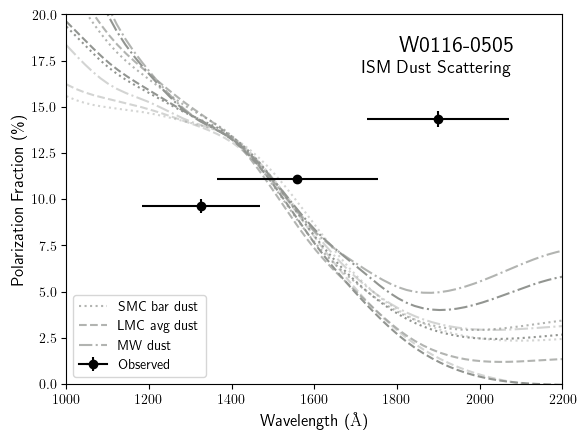}
    \caption{Polarization fraction as a function of wavelength for W0116--0505. The circles show the measured values, with the vertical error-bar showing the uncertainty and the horizontal error-bar showing the width of the broad-band filter used. The lines show the different model polarization fractions obtained by shining the quasar light into the ISM after optimizing both the torus half-opening angle $\psi$ and the inclination angle $\eta$ assuming the dust mixtures of \citet{draine03} for the SMC bar (dotted), for the LMC average (dashed) and for the Milky Way (dot-dashed). The heavier lines show the polarization obtained from simultaneously having backward and forward scattering, while the intermediate and lighter weight lines respectively show that for forward scattering and backward scattering only.}
    \label{fg:W0116_ISM_Dust}
\end{figure}

\begin{figure}
    \centering
    \includegraphics[width=0.49\textwidth]{"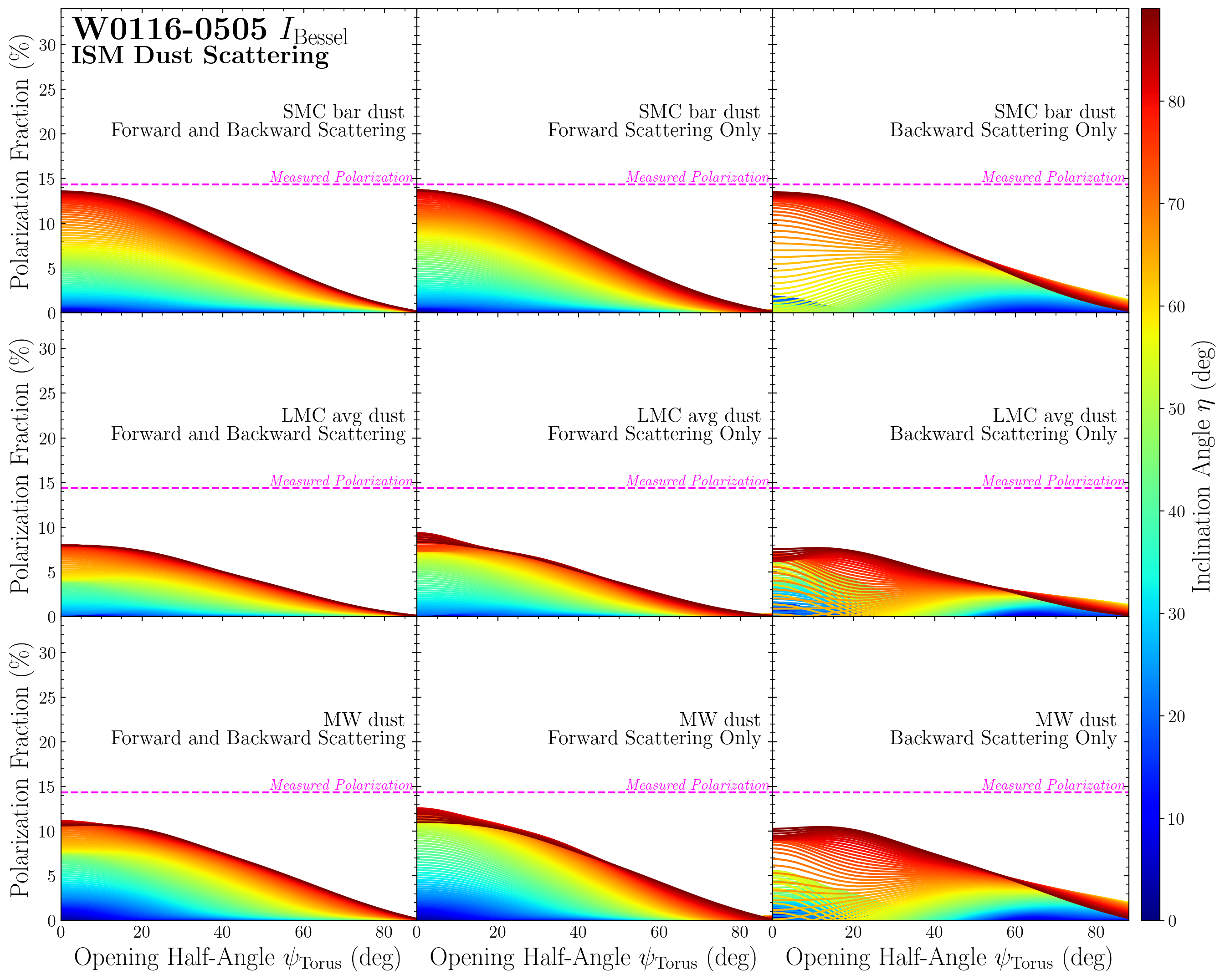"}
    \caption{Predictions for the polarization fraction expected in the $I_{\rm Bessel}$ band for W0116--0505 from an AGN with a torus half-opening angle of $\psi_{\rm Torus}$ shining on a diffuse dust component in the ISM. The color indicates the inclination of the system with respect to the line of sight, $\eta$. We show the predictions for the three dust mixtures of \citet{draine03} discussed in Section \ref{ssec:ism_scatt}, namely that of the SMC bar (top), the LMC average (middle) and the MW (bottom). For each dust mixture we show the expectations for a combination of forward and backward scattering (left), only for forward scattering (middle) and only for backward scattering (right). The magenta dashed line shows the measured polarization fraction in the $I_{\rm Bessel}$ band for W0116--0505.}
    \label{fg:ISM_W0116_Ibess}
\end{figure}

In the case of W0116--0505, we find that no set of parameters is able to reproduce the wavelength dependence of the polarization fraction. As shown in Figure \ref{fg:W0116_ISM_Dust}, all the considered dust mixtures have significantly higher polarization fractions toward bluer wavelengths while the measurements show the opposite. If we only focus on the $I_{\rm Bessel}$ band polarization, as it is the only band free from line contamination, we find that none of the models are able to fully reproduce the observed polarization fraction, as shown in Figure \ref{fg:ISM_W0116_Ibess}. We find that SMC-type dust gets closest to the observed value if the inclination $\eta$ is approximately 90~\rm deg and the torus half-angle $\psi_{\rm Torus}$ is small, $\lesssim 10~\rm deg$. Assuming $\psi_{\rm Torus}\lesssim 10~\rm deg$, the fraction of scattered light would require a column density of $N_{\rm H}\gtrsim 10^{22}~\rm cm^{-2}$ or $E(B-V)\gtrsim 0.15~\rm mag$ in the ISM of the host galaxy assuming the median dust-to-gas ratio of \citet{maiolino01} for dust in the circumnuclear regions of nearby AGN, which would be somewhat inconsistent with the lack of reddening found for the scattered light component by the SED model (see \S\ref{ssec:phot} and Table \ref{tab:targs}). The standard Galactic dust-to-gas ratio of \citet{guver09} is about 10 times larger than the median of \citet{maiolino01} and may be more appropriate for these dust mixtures, yet it would increase the tension with the value estimated from the SED modeling.

\begin{figure}
    \centering
    \includegraphics[width=0.49\textwidth]{"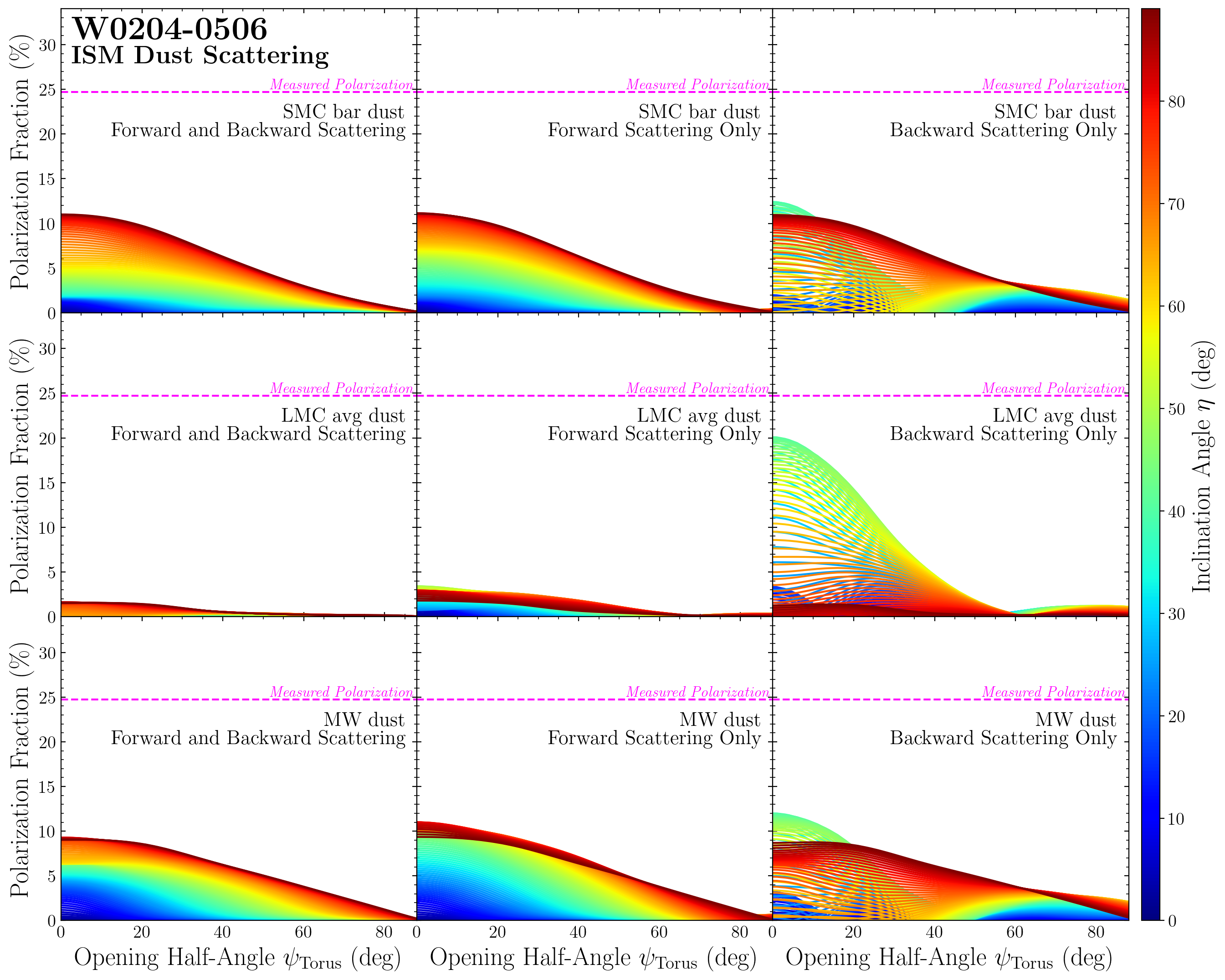"}
    \caption{Same as in Fig. \ref{fg:ISM_W0116_Ibess} but for the $R_{\rm Special}$ band in W0204--0506.}
    \label{fg:ISM_W0204_Rspec}
\end{figure}

The disagreement is much starker in the case of W0204--0506, as the highest polarization fraction achievable in the $R_{\rm Special}$ band by these dust mixtures is about $\sim10$\% for the SMC and MW dust mixtures (see Figure \ref{fg:ISM_W0204_Rspec}), far below the measured value of $24.72\pm 0.66\%$. The disagreement is discussed already by \citetalias{assef25}. They point out that LMC dust grains have a strong negative\footnote{We adopt here the usual convention of negative polarization implying a 90~deg shift in the polarization angle.} polarization at this wavelength \citep[see also discussion in][]{zakamska23} that could potentially help explain a higher polarization fraction. This is in fact the reason Figure \ref{fg:ISM_W0204_Rspec} shows such a small polarization fraction in the context of our model for W0204--0506 in the $R_{\rm Special}$ band for the combined forward and backward scattering model, as regions producing back-scattering cancel out part of the polarization fraction in regions producing front-scattering. Only allowing for back-scattering on LMC-type dust grains can raise the polarization up to about 20\% for an inclination $\eta$ of about $40~\rm deg$ and a small $\psi_{\rm Torus}\lesssim 10~\rm deg$. As back-scattering from LMC-type grains also has a significantly lower cross-section than front-scattering, the required column densities are quite high. For $\psi_{\rm Torus}\lesssim 10~\rm deg$, an $N_{\rm H}\gtrsim 8\times 10^{22}~\rm cm^{-2}$ is required to account for the 1.1\% scattered light fraction in this target, which translates to an $E(B-V)\gtrsim 1.2~\rm mag$ assuming the median dust-to-gas ratio of \citet{maiolino01} or to $E(B-V)\gtrsim 12~\rm mag$ assuming the Galactic standard. Both are inconsistent with the obscuration of 0.1~\rm mag for the scattered light found in \S\ref{ssec:phot}. 

Hence, even if we go to the most extreme conditions of this model where the obscured quasar shines onto an optically thin ISM, we are not able to reproduce the polarization fractions and other properties of W0116--0505 and W0204--0506. Changes to the dust properties could help, but it is difficult to expect very different dust mixtures in the host galaxies unless they are driven by the quasar itself. In section \ref{ssec:scat_polar_outflow} we discuss the possibility that the scattering medium corresponds to a dusty quasar outflow which can have very different dust mixtures than the standard ISM of galaxy, and we find a much better agreement with the observations. 

\end{appendix}

\end{document}